\documentclass[%
twocolumn,
superscriptaddress,
 amsmath,amssymb,
 prx,
showkeys,
showpacs,
]{revtex4-1}

\usepackage{graphicx}
\usepackage{dcolumn}
\usepackage{bm}
\usepackage{siunitx}
\usepackage{braket}
\usepackage{hyperref}
\usepackage{xcolor}
\usepackage{multirow}
\usepackage{array}
\usepackage{float}
\usepackage[title]{appendix}
\usepackage{etoolbox, chngcntr}
\usepackage[capitalize]{cleveref}
\usepackage{xr}

\makeatletter
\def\maketitle{
\@author@finish
\title@column\titleblock@produce
\suppressfloats[t]}
\makeatother

\newcommand{\circumeq}{\mathrel{\widehat{=}}} 

\floatstyle{plaintop}
\restylefloat{table}

\begin{document}
\begin{abstract}
    
\end{abstract}
\title{Long-distance transmon coupler with CZ gate fidelity above \SI{99.8}{\percent}}
\author{Fabian Marxer}
\thanks{These two authors contributed equally to this work}

\author{Antti Veps\"al\"ainen}
\thanks{These two authors contributed equally to this work}
\email{avepsalainen@meetiqm.com}

\author{Shan W. Jolin}

\author{Jani Tuorila}

\author{Alessandro Landra}

\author{Caspar Ockeloen-Korppi}

\author{Wei Liu}

\author{Olli Ahonen}
\affiliation{IQM Quantum Computers, Espoo 02150, Finland}

\author{Adrian Auer}
\affiliation{IQM Quantum Computers, Munich 80636, Germany}

\author{Lucien Belzane}

\author{Ville Bergholm}

\author{Chun Fai Chan}

\author{Kok Wai Chan}

\author{Tuukka Hiltunen}

\author{Juho Hotari}

\author{Eric Hyypp\"a}

\author{Joni Ikonen}

\author{David Janzso}

\author{Miikka Koistinen}

\author{Janne Kotilahti}

\author{Tianyi Li}

\author{Jyrgen Luus}
\affiliation{IQM Quantum Computers, Espoo 02150, Finland}

\author{Miha Papic}
\affiliation{IQM Quantum Computers, Munich 80636, Germany}

\author{Matti Partanen}

\author{Jukka R\"abin\"a}

\author{Jari Rosti}

\author{Mykhailo Savytskyi}

\author{Marko Sepp\"al\"a}

\author{Vasilii Sevriuk}

\author{Eelis Takala}

\author{Brian Tarasinski}
\affiliation{IQM Quantum Computers, Espoo 02150, Finland}

\author{Manish J. Thapa}
\affiliation{IQM Quantum Computers, Munich 80636, Germany}

\author{Francesca Tosto}

\author{Natalia Vorobeva}

\author{Liuqi Yu}

\author{Kuan Yen Tan}

\author{Juha Hassel}
\affiliation{IQM Quantum Computers, Espoo 02150, Finland}

\author{Mikko M\"ott\"onen}
\affiliation{IQM Quantum Computers, Espoo 02150, Finland}
\affiliation{QCD Labs, QTF Centre of Excellence, Department of Applied Physics, Aalto University, Aalto 00076, Finland}
\affiliation{VTT Technical Research Centre of Finland, QTF Center of Excellence, VTT, VTT 02044, Finland}

\author{Johannes Heinsoo}
\affiliation{IQM Quantum Computers, Espoo 02150, Finland}

\begin{abstract}
Tunable coupling of superconducting qubits has been widely studied due to its importance for isolated gate operations in scalable quantum processor architectures. Here, we demonstrate a tunable qubit-qubit coupler based on a floating transmon device which allows us to place qubits at least \SI{2}{\milli\meter} apart from each other while maintaining over $\SI{50}{\mega\hertz}$ coupling between the coupler and the qubits. In the introduced tunable-coupler design, both the qubit-qubit and the qubit-coupler couplings are mediated by two waveguides instead of relying on direct capacitive couplings between the components, reducing the impact of the qubit-qubit distance on the couplings. This leaves space for each qubit to have an individual readout resonator and a Purcell filter needed for fast high-fidelity readout. In addition, simulations show that the large qubit-qubit distance significantly lowers unwanted non-nearest-neighbor coupling and allows multiple control lines to cross over the structure with minimal crosstalk. Using the proposed flexible and scalable architecture, we demonstrate a controlled-$Z$ gate with \SI[separate-uncertainty = true]{99.81(2)}{\percent} fidelity.
\end{abstract}
\date{\today}
\maketitle

\section{Introduction}

\begin{figure*}[tb]
    \centering
    \includegraphics[width=1.0\textwidth]{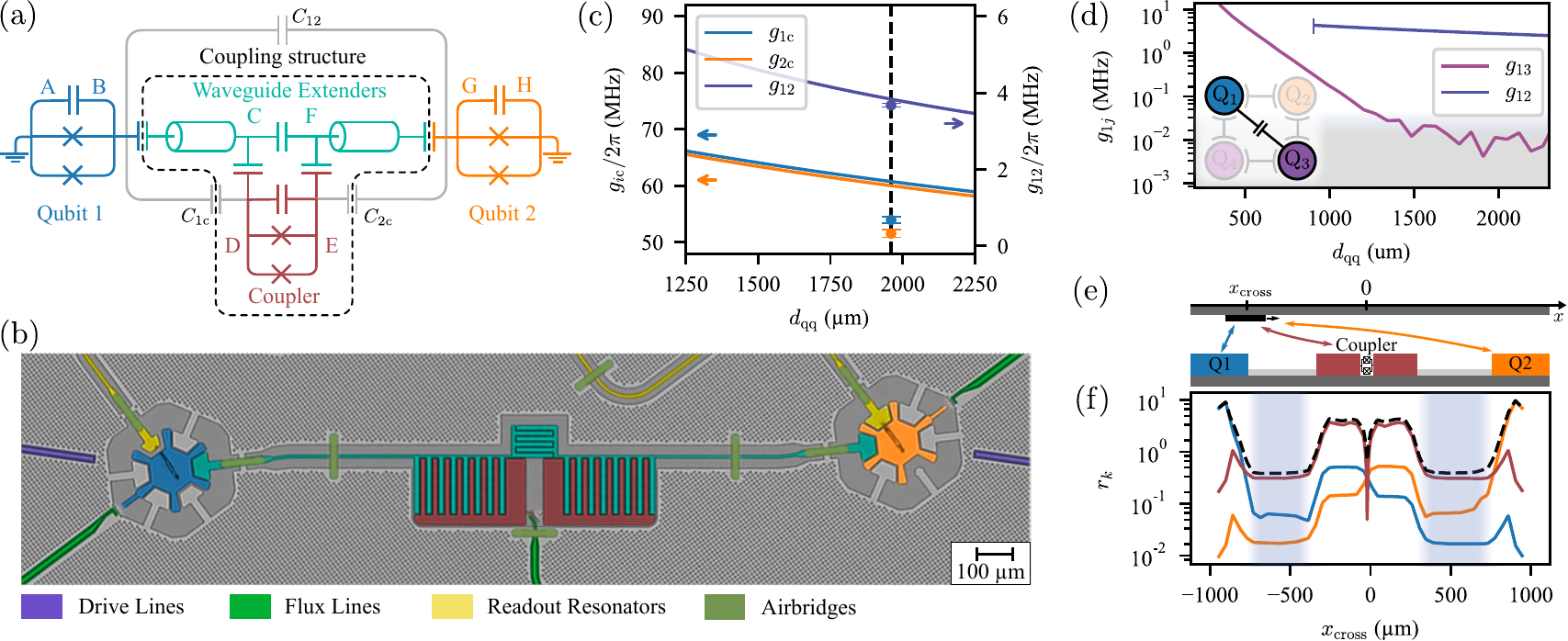}
    \caption{(a) Quasi-lumped-element circuit diagram of qubits (blue and orange) coupled by a tunable coupling structure consisting of waveguide extenders (turquoise) and a floating coupler qubit (red). The gray lines show the effective lumped-element model for the coupling capacitances arising from the black waveguide extenders C and F.
    (b) False-colored micrograph of the qubit-coupler-qubit system. The qubits, coupler, and waveguide extenders follow the color code from (a). The airbridges are colored non-transparently to protect unpublished intellectual property.
    (c) Simulation of the effective coupling strengths between qubit 1 and the coupler, $g_{\rm 1c}$, qubit 2 and the coupler, $g_{\rm 2c}$, and qubit 1 and qubit 2, $g_{\rm 12}$, for different qubit-qubit distances $d_{\rm qq}$ (solid lines). The measured coupling values (dots with \SI{68}{\percent} confidence intervals) are shown for our device with $d_{\rm qq} = \SI{1960}{\micro\meter}$ (dashed line).
    (d) Simulated spurious next-nearest-neighbor coupling $g_{13}$ between $\rm{Q}_1$ and $\rm{Q}_3$ in a square lattice configuration (schematic), and the nearest-neighbor coupling $g_{12}$ as a function of the distances $d_{\rm qq}$ between $\rm{Q}_1$ and $\rm{Q}_2$. The gray shaded area indicates the region, where the coupling values fall below the simulation accuracy.
    (e) Schematic of a simplified flip-chip architecture, where the qubits and the coupler are connected via waveguide extenders (light gray) on the bottom chip and a long perpendicular transmission line (black) is crossing above the qubit-coupler structure on the top chip.
    (f) Simulated capacitance coupling ratios (see main text) from each component to the transmission line (color code same as in (e)) at various crossing positions $x_{\rm cross}$ of the transmission line, total coupling (black dashed line) and region for low-crosstalk crossings (blue shading).
    }
    \label{fig:fig_1}
\end{figure*}

The implementation of high-fidelity two-qubit gates is a key requirement for scalable quantum processors \cite{reagor2018demonstration, arute2019quantum, chen2021exponential}. The performance of quantum gates relying only on the static qubit-qubit coupling \cite{majer2007coupling, dicarlo2009demonstration, kandala2019error} is typically limited by stray $ZZ$ interaction between the qubits resulting in long gate times due to a poor on/off ratio for the coupling. To address this problem, different tunable-coupler designs have been proposed \cite{hime2006solid, niskanen2007quantum, chen2014qubit, geller2015tunable, chen2022tuning} with increasing sophistication.

An important step toward tunable couplers with a high on/off ratio and with minimal impact on qubit coherence was the observation that one can design a coupler-mediated tunable interaction that cancels out the static qubit-qubit coupling at a specific coupler off-frequency which is above the qubit frequencies~\cite{yan2018tunable}. Since then, such couplers have been successfully used in several experiments \cite{arute2019quantum, li2020tunable, xu2020high, collodo2020implementation, sung2021realization, ye2021realization}. In Ref.~\cite{sete2021floating}, a floating transmon coupler that can also be operated below the qubit frequencies was proposed, allowing two-qubit-gate operations near the flux insensitive coupler sweet spot and thereby reducing the impact of coupler flux noise during the gate. Moreover, the concept was used in Ref.~\cite{stehlik2021tunable} for implementing high-fidelity CZ gates for fixed-frequency floating transmons.

Although the $ZZ$-interaction-free tunable-coupler designs have seen significant success in enabling high-fidelity two-qubit gates, the static qubit-qubit coupling needed for canceling the coupler-mediated interaction arises from the direct coupling between the qubits, controlled mainly by the qubit-qubit distance. Having no residual $ZZ$ coupling in such architectures therefore requires small qubit-qubit distances. This restriction introduces severe limitations for the qubit locations, leading to high spurious non-nearest-neighbor coupling, and furthermore, limits the space in a square qubit lattice, allowing only the most essential components to be placed in between the lattice sites.

Here, we introduce and experimentally demonstrate an extended floating coupler that allows us to increase the physical distances between qubits, thus providing the space on the chip for readout resonators with individual Purcell filters for high-fidelity readout \cite{heinsoo2018rapid, Sete2015}. Furthermore, long qubit-qubit distances significantly lower the parasitic non-nearest-neighbor coupling and enable low-crosstalk paths for multiple control lines above the qubit-coupler-qubit structure if flip-chip technology is utilized. Similar to Refs.~\cite{sete2021floating, stehlik2021tunable}, the tunable coupler is operated below the qubit frequencies, allowing two-qubit-gate operations closer to its flux insensitive sweet spot, and furthermore, preventing the coupler to cross any qubit readout-resonator modes. The ability to have long qubit-qubit distances is achieved using two waveguide extenders that mediate the direct qubit-qubit coupling, as well as the couplings between the qubits and the coupler. These couplings predominantly originate from interdigital capacitors between the two waveguide extenders and the coupler, making it possible to maintain couplings with similar magnitude as previous designs \cite{sung2021realization, li2020tunable} across a broad range of qubit-qubit distances. To demonstrate that the couplings between the qubits and the coupler are high enough for qubit-qubit distances beyond \SI{1}{\milli \meter}, we implement a fast and high-fidelity controlled-$Z$ (CZ) gate with a duration of \SI{33}{\nano\second} and \SI[separate-uncertainty = true]{99.81(2)}{\percent} fidelity for a qubit-qubit distance of \SI{1.96}{\milli\meter}, which is four times longer than in typical tunable coupler designs~\cite{sung2021realization, collodo2020implementation}. Our coupler design is readily extendable to a square qubit lattice, making it an appealing building block for a scalable high-fidelity quantum processor.

\section{Extended floating transmon coupler design}

We propose a coupler design in which two grounded transmon qubits interact through a floating transmon coupler. As shown in Fig.~\ref{fig:fig_1}(a), the qubits are connected to the coupler with two waveguide extenders that mediate effective capacitances $C_{\rm 1c}$, $C_{\rm 2c}$, and $C_{12}$ between the qubits and the coupler. The two waveguide extenders shown in Fig.~\ref{fig:fig_1}(b) are implemented as \SI{0.9}{\milli\meter}-long coplanar waveguides whose resonance frequencies are an order of magnitude higher than the qubits and coupler frequencies, such that the waveguide extender modes do not couple to the qubit and coupler modes.
At one end, the waveguide extenders are coupled to each other and to the coupler via interdigital capacitors, and at the other end to the qubits via gap capacitors. These gap capacitors are designed to accommodate four waveguide extenders which are coupled to a single qubit to form a square lattice. All the capacitance values between the extenders, the qubits, and the coupler can be independently adjusted to realize the desired effective capacitances, see Appendix~\ref{sec:effective_coupling} for the detailed model. 
Within the effective capacitance model, the dynamics of the system can be described using the coupling strengths between qubit 1 ($\rm{Q}_1$) and the coupler, $g_{\rm 1c}$, qubit 2 ($\rm{Q}_2$) and the coupler, $g_{\rm 2c}$, and the qubit-qubit coupling $g_{\rm 12}$.

In contrast to direct capacitive couplings between the qubits and the coupler studied in Ref.~\cite{sete2021floating}, the waveguide extenders facilitate flexible positioning of the individual components by simply elongating the waveguide extenders. To investigate the impact of qubit-qubit distance in our system, we calculate the effective coupling strengths for waveguide extenders of varying lengths using simulated capacitance values between all the superconducting islands, A--H in Fig.~\ref{fig:fig_1}(a). As shown in Fig.~\ref{fig:fig_1}(c), changing the distance $d_{\rm qq}$ between the qubits by up to \SI{1000}{\micro\meter} impacts the coupling strengths between the qubits and the coupler by less than \SI{50}{\percent}, implying that the couplings are still large enough for fast two-qubit gates over a wide range of qubit-qubit distances. In the simulation, we use scikit-rf \cite{scikit} to model the qubits and the coupler as lumped-element circuits, and apply the transmission line model for the waveguide extenders, see Appendix~\ref{sec:simulation_coupling_strength} for details on the simulation. The slight decrease in coupling strengths with increasing $d_{\rm qq}$ can be attributed to the increasing extender capacitance to ground which starts to contribute to the total capacitance of the system, ultimately limiting $d_{\rm qq}$.

Even though the test device studied in this work has only two qubits connected using the coupler, the layout is designed such that identical couplings can be simultaneously reached for up to four neighboring qubits, compatible with scaling to large qubit lattices. In such a lattice, the resulting freedom of qubit spacing allows the spurious couplings to non-nearest-neighbor qubits to be reduced while maintaining strong nearest-neighbor (NN) coupling, which is essential for a $ZZ$-interaction-free idling configuration of a multi-qubit system. The dominating non-nearest-neighbor couplings in a square qubit lattice architecture are the diagonal next-nearest-neighbor (NNN) qubit pairs. As shown in Fig.~\ref{fig:fig_1}(d), the simulated spurious NNN coupling $g_{13}$ between two diagonally placed qubits with the typical qubit-qubit distance of $d_{\rm qq}= \SI{500}{\micro\meter}$ has the same order of magnitude as the typical NN coupling $g_{12}$ \cite{sung2021realization, li2020tunable}, potentially leading to large residual $ZZ$-couplings during idling time. As $g_{13}$ decreases with increasing $d_{\rm qq}$, the NNN coupling at our qubit-qubit distance of $\SI{1960}{\micro\meter}$ is less than $g_{13}<\SI{30}{\kilo\hertz}$, which is at least two orders of magnitude smaller than $g_{12}$. The upper bound of the estimate is limited by the accuracy of the simulation. The NNN coupling $g_{13}$ was extracted from an Ansys HFSS finite-element-method (FEM) simulation of the direct capacitance between qubits ${\rm Q_1}$ and ${\rm Q_3}$, and did not include other qubits or couplers. However, we verified that the inclusion of $\rm{Q}_2$ and the couplers connecting $\rm{Q}_1$, $\rm{Q}_2$, and $\rm{Q}_3$ in the simulation did not show a significant difference. From these capacitances, we calculate $g_{13}$ by modeling the two qubits as coupled harmonic oscillators with a frequency of $\SI{4.3}{\giga\hertz}$ \cite{yan2018tunable}. For the simulation of the mediated qubit-qubit coupling $g_{12}$ the coupler was included, limiting the smallest simulation range of the qubit-qubit distance to $d_{\rm qq}=\SI{920}{\micro\meter}$ corresponding to the width of the coupler.

In addition to suppressing the NNN coupling, the large qubit-qubit distance gives us physical space for readout structures with individual Purcell filters for each qubit \cite{Sete2015, heinsoo2018rapid}, enabling fast and high-fidelity readout \cite{Walter2017rapid}. Implementing e.g.~\SI{6}{\giga\hertz} readout resonators and Purcell filters as spiral $\lambda / 4$ coplanar resonators requires together at least \SI{1}{\milli\meter^2} of space, which easily fits in a square qubit lattice with our qubit-qubit distances of $d_{\rm qq} = \SI{1960}{\micro\meter}$.

\begin{figure*}[tb]
    \centering
    \includegraphics[width=1.0\textwidth]{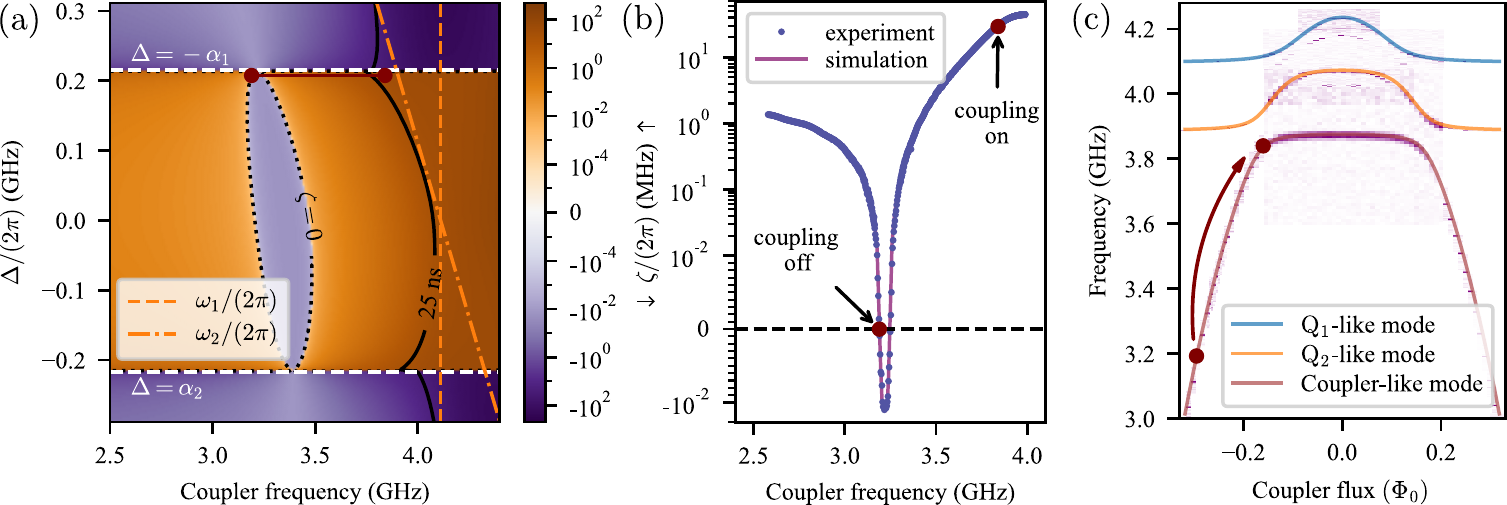}
    \caption{(a) The effective $ZZ$ interaction strength $\zeta$ between the qubits is shown as a function of the coupler frequency $\omega_{\rm c}/(2\pi)$ and the frequency detuning between the two qubits, $\Delta/(2\pi)$. If $\alpha_2 < \Delta < -\alpha_1$, there exists a contour (black dotted line) along which the $ZZ$ interaction vanishes.
    The qubit frequencies are shown with orange dashed lines and the detuning values corresponding to their anharmonicities are marked with white dashed lines. The solid black line indicates the lowest coupler frequency at which a conditional phase of $\phi_{11} = \pi$ can be accumulated in \SI{25}{\nano\second}.
    The solid red line illustrates the path along which the coupler is operated in the experiments. 
    (b) Experimentally measured interaction strength (dots) as a function of the coupler frequency $\omega_{\rm c} / (2\pi)$ in units of flux quanta $\Phi_0$ for the specific qubit-qubit detuning chosen to implement the CZ gate. The numerical model (purple line) is fitted to the data to extract the coupling strengths between the two qubits and between each qubit and the coupler.
    (c) Experimentally measured eigenfrequencies (white to blue spectrum) and fitted eigenfrequencies (lines) of the qubits and the coupler as a function of the external flux $\Phi_{\rm ext}^{\rm c}$ threading through the coupler SQUID loop. The coupler spectrum below \SI{3.7}{\giga\hertz} is measured using ${\rm Q_2}$ as an ancilla, see Appendix~\ref{sec:coupler_characterization}. The red circles show the idling and the operation configuration of the coupler. 
    }
    \label{fig:fig_2}
\end{figure*}

An additional benefit of the waveguide extenders is that the electric field density of the qubit and the coupler modes is reduced above the extenders. This allows microwave control lines to cross the qubit-coupler structure with low crosstalk. To support this claim, we conduct a FEM simulation to extract the capacitance between the components on the chip and a transmission line ($\rm TL$) that crosses the structure \SI{8}{\micro\meter} above the qubits (Fig.~\ref{fig:fig_1}(e)), corresponding to a typical flip-chip topology \cite{kosen2022building}, see Appendix~\ref{sec:simulation_crosstalk} for further details on the simulation. To put the capacitance values into context, we calculate the drive crosstalk from the capacitance ratio $r_k = C_{{\rm k} \leftrightarrow \rm{TL}}/C_{\rm Q \leftrightarrow \rm{DL}}$, where $C_{{\rm k} \leftrightarrow \rm{TL}}$ is the capacitance between the center conductor of the transmission line and component $k = \{{\rm Q_1}, {\rm Q_2}, {\rm c}\}$, and $C_{\rm Q  \leftrightarrow \rm{DL}}=\SI{0.12}{\femto\farad}$ is the capacitance between each qubit and its corresponding drive line, see Fig.~\ref{fig:fig_1}(f). For this geometry, crossing the tunable coupling structure above the waveguide extenders reduces the parasitic coupling by a factor of 10 compared to passing over the tunable coupler. Above the extenders, the crosstalk is $r_{\rm Q_i} < \num{0.1} = \SI{-20}{\dB}$, which is comparable to the typical crosstalk between two microwave control lines and can be further reduced by increasing the chip-to-chip distance. This level of crosstalk enables the implementation of qubit lattices with crossing lines without resorting to more sophisticated technologies such as through silicon vias, significantly reducing the complexity of the scalable architecture.

\section{CZ gate implementation}

In our design, the effective static $ZZ$ interaction strength $\zeta = \omega_{11} - \omega_{10} - \omega_{01} + \omega_{00}$ between the qubits can be eliminated if the detuning between the two qubits is smaller than their anharmonicity \cite{sete2021floating}, $\Delta = \omega_1 - \omega_2 \in [\alpha_2, -\alpha_1]$, where $\alpha_i$ and $\omega_i$ are the anharmonicity and the angular frequency of qubit $i$ in the laboratory frame, and $\omega_{ij}/ (2\pi)$ is the eigenfrequency for the computational state in which $\rm{Q}_1$ is in state $i$ and $\rm{Q}_2$ is in state $j$. If this condition is met, a contour with $\zeta = 0$ for coupler frequencies below the qubit frequencies can be identified, as depicted by the simulation shown in Fig.~\ref{fig:fig_2}(a). In the simulation, we numerically solve for the $ZZ$ interaction strength in the dressed-state basis comprising the three lowest-energy states of the coupler and the qubits, see Appendix~\ref{sec:effective_coupling} for details.

The $ZZ$ interaction can be turned on by tuning the coupler frequency away from the $\zeta=0$ contour. The highest coupling and therefore the fastest gate can be reached when the coupler frequency is close to the qubit frequencies, as shown in the \SI{25}{\nano\second}-gate-time curve in Fig.~\ref{fig:fig_2}(a). The most relevant region for a fast and high-fidelity CZ gate is close to $\Delta = -\alpha_1$, where the qubit states evolve through a near-resonant $\ket{11} \leftrightarrow \ket{20}$ transition if the coupling is turned on, implementing the diabatic version of the CZ gate. In the experiment, we set the qubit frequencies at $\omega_1 = \omega_2 - a \alpha_1$, where $\omega_1/(2\pi) = \SI{4.10}{\giga\hertz}$, $\omega_2/(2\pi) = \SI{3.89}{\giga\hertz}$, $\alpha_1/(2\pi) = \SI{-216}{\mega\hertz}$ and $a = 0.97$, see Appendix~\ref{sec:experimental_setup} for the full experimental setup and Appendix~\ref{sec:sample_fabrication} for details on sample fabrication. Even though the fastest gate times can be reached when the $\ket{11}$ and $\ket{20}$ transitions are resonant, i.e. $a = 1$, we deliberately operate the qubits slightly away from the resonance to mitigate the effect of drive crosstalk during single-qubit gates. The drive crosstalk could also be reduced by idling the qubits in a configuration where the states $\ket{11}$ and $\ket{20}$ are initially far-detuned and tuning the qubits to their operation point only during the CZ gate \cite{sung2021realization}. However, in order to reduce the number of required microwave control lines in the system, we choose not to use fast flux control for the qubits.

\begin{figure}[tb]
    \centering
    \includegraphics[width=1.0\columnwidth]{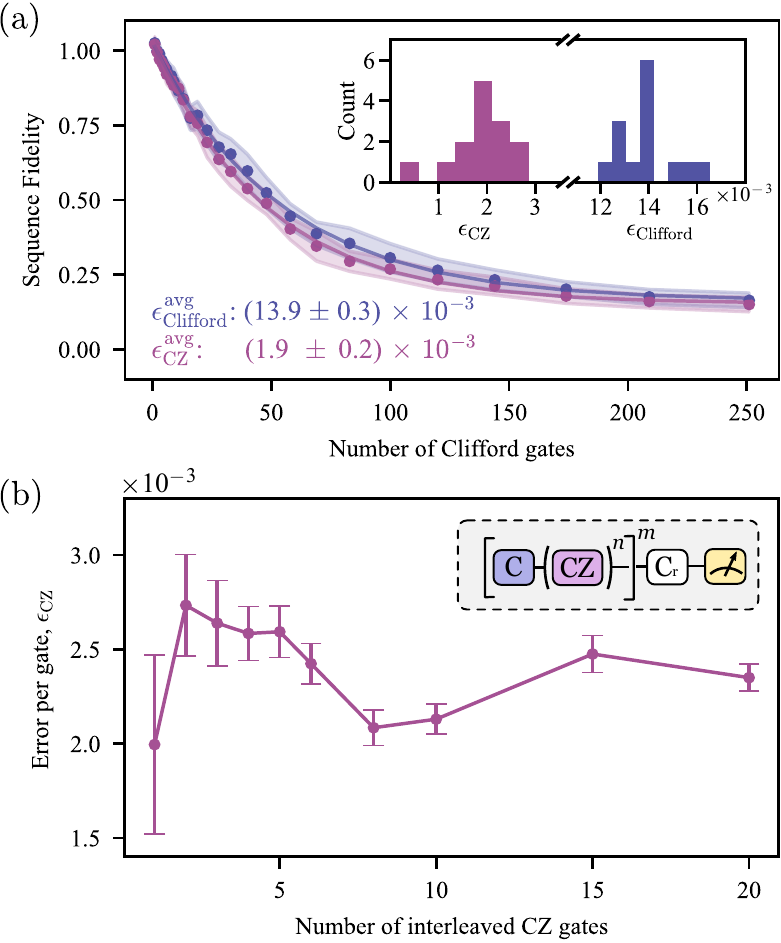}
    \caption{Interleaved randomized benchmarking experiment for two qubits. (a) Experimental data of the sequence fidelity for randomized benchmarking with interleaved CZ gates (purple dots) and the reference Clifford sequence (blue dots) as a function of the number of Clifford gates. The purple and blue shaded regions indicate the standard deviation of 30 random sequences for the interleaved randomized benchmarking experiment and the reference Clifford sequence, respectively. The histograms in the inset show the error per Clifford gate (blue) and the error per CZ gate (purple) for the 14 repeated experiments, from which the average error per Clifford gate of $\epsilon^{\rm{avg}}_{\rm{Clifford}} = \num[separate-uncertainty]{1.39(3)e-2}$ and the average error per CZ gate of $\epsilon^{\rm{avg}}_{\rm{CZ}} = \num[separate-uncertainty]{1.9(2)e-3}$ were estimated. 
    (b) Experimentally obtained error per CZ gate as a function of the number of interleaved CZ gates $n$ in the randomized benchmarking sequence. The error bars indicate the standard deviation extracted from the residuals of the best model fit of the sequence decay for $n$ interleaved CZ gates. The gate sequence is depicted in the inset: A random Clifford gate $\rm C$ followed by $n$ CZ gates, repeated $m$ times, after which a reversing Clifford gate $\rm C_r$ is applied. Extracting the fitting error for the sequence fidelity for $n$ interleaved CZ gates yields a reduction in the uncertainty of the error per CZ gate by a factor of $1/n$.}
    \label{fig:fig_3}
\end{figure}

We experimentally determine the coupler idling frequency by measuring the conditional phase of $\rm{Q}_1$ when $\rm{Q}_2$ is initialized either in the ground or in the excited state, and find $\lvert \zeta \rvert <\SI{2}{\kilo\hertz}$ at $\omega_{\rm c}/(2\pi) = \SI{3.195}{\giga\hertz}$, see Fig. \ref{fig:fig_2}(b). The fit to the measured  $ZZ$ interaction strength as a function of the coupler frequency gives an estimate of $g_{\rm 1c}/(2\pi) = \SI[separate-uncertainty]{51.5(7)}{\mega\hertz}$, $g_{\rm 2c}/(2\pi) = \SI[separate-uncertainty]{53.9(6)}{\mega\hertz}$, and $g_{\rm 12}/(2\pi) = \SI[separate-uncertainty]{3.7(5)}{\mega\hertz}$ where the couplings are defined at the idling configuration. The fitted coupling strengths deviate less than $\SI{15}{\percent}$ from the design and provide the strong $ZZ$ interaction strength needed to implement a CZ gate in less than \SI{25}{\nano\second} interaction time. In this configuration, we measure simultaneous single-qubit gate errors $\epsilon_{\rm Q1}^{\rm sim} = \num[separate-uncertainty]{1.32(5)e-3}$ and $\epsilon_{\rm Q2}^{\rm sim} = \num[separate-uncertainty]{8.9(5)e-4}$, which are only slightly higher than individual single-qubit gate errors, $\epsilon_{\rm Q1} = \num[separate-uncertainty]{7.8(3)e-4}$ and $\epsilon_{\rm Q2} = \num[separate-uncertainty]{6.8(4)e-4}$, respectively.

To realize the diabatic version of the CZ gate \cite{barends2019diabatic}, we turn on the effective exchange interaction between the states $\ket{20}$ and $\ket{11}$ of the qubits by applying a \SI{22}{\nano\second}-long flux pulse to the coupler, shifting its frequency from the idling point to the operation point, see Fig.~\ref{fig:fig_2}(c). We use a Slepian-shaped flux pulse to minimize the dominating leakage processes during the gate from $\rm{Q}_2$ to the coupler \cite{martinis2014fast, sung2021realization}. In addition, to account for the distortions in the flux pulse shape, we use two consecutive infinite-impulse-response filter to correct for their shape in real-time \cite{rol2020time}. To mitigate the impact of the flux distortion at the nanosecond time scale, we add a \SI{5.5}{\nano\second} idle time before and after the flux pulse, resulting in a total flux pulse duration of \SI{33}{\nano\second}.

\begin{figure*}[htb]
    \centering
    \includegraphics[width=1.5\columnwidth]{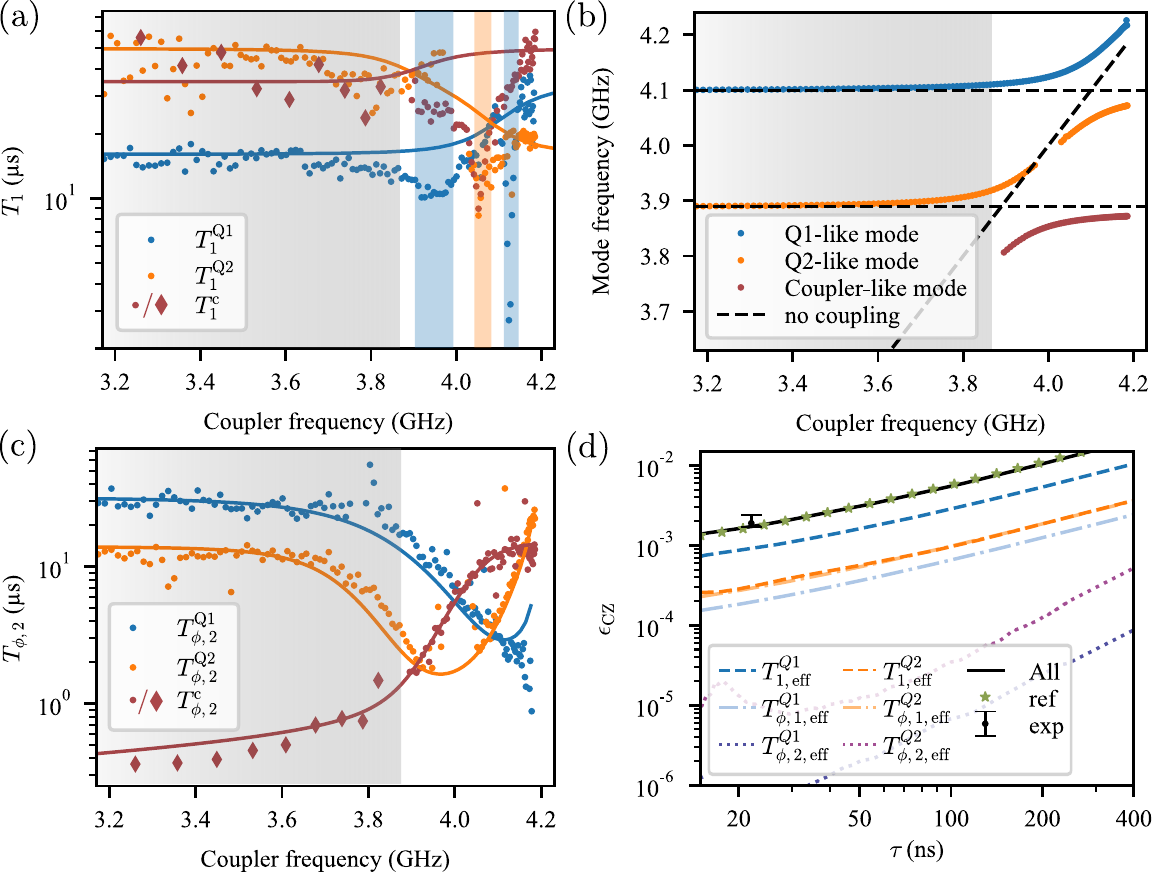}
    \caption{The impact of the coupler decoherence on the qubits.
    (a) Energy relaxation times of the hybridized qubit-coupler modes measured directly (dots), measured via an ancilla qubit (diamonds), and evaluated numerically based on the hybridization with the coupler (solid lines) as a function of the coupler frequency $\omega_{\rm c}/(2\pi)$. The shaded areas indicate the coupler frequencies swept during the CZ gate (gray) and regions with unknown sources of decoherence, partly attributed to parasitic two-level systems (orange and blue shading for $\rm{Q}_1$ and $\rm{Q}_2$, respectively).
    (b) Measured frequencies of hybridized modes during coherence measurements (dense set of dots) and frequencies of the uncoupled system (black dashed lines). 
    (c) as in (a) but shows the Gaussian component of the pure dephasing time for the hybridized qubit and coupler modes.
    (d) Coherence limit of the CZ gate (solid black line) as a function of the interaction time $\tau$ calculated using Eq.~\eqref{eq:coherence_limit} with contributions from individual coherence times (dashed lines), and the coherence limit for no decoherence contributions from the coupler (green stars). When evaluating the gate error, an idling time of \SI{11}{\nano\second} is added on top of the interaction time. The black dot shows the experimental gate error with the \SI{68}{\percent} confidence interval. The data for $T_{\phi, 1, {\rm eff}}^{\rm Q_2}$ is below the curve corresponding to $T_{1, {\rm eff}}^{\rm Q_2}$ and not well visible in the plot.
    }
    \label{fig:fig_4}
\end{figure*}

We optimize the CZ gate by adjusting the amplitude and duration of the coupler flux pulse to find an operation point that minimizes the occupation in the state $\ket{20}$ while accumulating a conditional phase shift $\phi_{11} = \pi$ in state $\ket{11}$. We account for the additional single-qubit phase accumulation caused by the dispersive interaction between the coupler and the qubits during the gate by applying virtual $Z$ gates \cite{mckay2017efficient} to both qubits after each CZ gate. 

\section{CZ gate characterization}

To characterize the average gate fidelity, we employ randomized benchmarking where a sequence of random two-qubit Clifford gates is interleaved with CZ gates \cite{magesan2012efficient, barends2014superconducting}. The experiment was repeated 14 times over a period of \SI{15}{\hour}, shown in Fig.~\ref{fig:fig_3}(a). We extract an average error per CZ gate of $\epsilon^{\rm avg}_{\rm CZ} = \num[separate-uncertainty]{1.9(2)e-3}$ from the collected data, where the uncertainty indicates the \SI{68}{\percent} confidence interval. For consistency, the average error per Clifford $\epsilon_{\rm Clifford} = \num[separate-uncertainty]{1.39(3)e-2}$ can be compared to the errors of its constituent CZ and single-qubit gates. On average, each  two-qubit Clifford gate consists of 1.5 CZ gates and 8.25 single-qubit gates \cite{barends2014superconducting}, which by simple summation of errors results in the estimate for the Clifford error $\epsilon_{\rm Clifford}^{\rm estimate} = \num[separate-uncertainty]{1.1(4)e-2}$, in agreement with the measured value.

We can improve the estimate for the CZ gate error by interleaving multiple CZ gates in the randomized benchmarking sequence and calculating the average error per gate for different numbers $n$ of interleaved gates \cite{sheldon2016characterizing}. As shown in Fig.~\ref{fig:fig_3}(b), the error per CZ gate converges from $\epsilon_{\rm CZ}^{(1)} = \num[separate-uncertainty]{2.0(5)e-3}$ for one interleaved CZ gate towards $\epsilon_{\rm CZ}^{(20)} = \num[separate-uncertainty]{2.35(7)e-3}$ for 20 interleaved CZ gates, demonstrating a seven-fold reduction in the uncertainty and a minor increase for the error per CZ gate, see Appendix~\ref{sec:ncz_irb} for further details. In contrast to conventional randomized benchmarking, where coherent errors are suppressed due to randomization of the basis between each application of the CZ gate, interleaving the sequence with several CZ gates reveals coherent errors as well since they tend to add up for the $n$ interleaved CZ gates. The minor increase in the error rate when interleaving multiple CZ gates suggests that the CZ gate fidelity is mostly limited by decoherence at its chosen operation point, rather than by coherent errors. The low error rate per CZ gate in the iterated randomized benchmarking experiment is important for real-world applications due to the accumulation of coherent errors being particularly detrimental to many quantum algorithms \cite{kjaergaard2022demonstration}.

During the CZ gate, the coupler and the qubits become strongly hybridized which affects the coherence of the computational states. To quantify the impact of decoherence during the CZ gate, we first measure the energy relaxation times of all the three hybridized qubit and coupler modes as a function of the uncoupled coupler frequency, shown in Fig.~\ref{fig:fig_4}(a).
Even though there are no individual drive and readout lines for the coupler, we can  measure the energy-relaxation time of the coupler-like mode using the qubit readout and drive lines when the coupler and qubit states are hybridized. For the coupler frequencies below \SI{3.9}{\giga\hertz} the degree of hybridization is low and we employ two-probe readout with $\rm{Q}_2$ as an ancilla qubit instead, see Appendix~\ref{sec:coupler_characterization} for details on the experiment. The measured frequencies of the hybridized modes are shown in Fig. \ref{fig:fig_4}(b). In the frequency range relevant for the CZ gate (gray shaded area), we observe that the energy-relaxation times of the qubit-like modes are not significantly affected by the hybridization with the coupler due to its high intrinsic energy-relaxation time, $T_1^{\rm c} > \SI{40}{\micro\second}$ at the idling configuration.

To model the dependency of the energy relaxation times of the hybridized states on the coupler frequency, we assume that the energy relaxation times of the uncoupled modes stay constant as a function of the coupler frequency. We then numerically evaluate the energy relaxation time of the coupled modes based solely on the degree of hybridization with the coupler, shown with solid lines in Fig~\ref{fig:fig_4}(a). In the frequency range relevant for the CZ gate and at coupler frequencies above \SI{4.1}{\giga\hertz}, we find a good match between the measurements and the model. For coupler frequencies between \SI{3.9}{\giga\hertz} and \SI{4.1}{\giga\hertz}, we observe unexpected dips in the relaxation times, which can potentially be attributed to parasitic two-level systems (TLSs) near ${\rm Q_1}$ and ${\rm Q_2}$ \cite{klimov2018fluctuations, mueller2019towards, krinner2022realizing}. However, these TLSs do not have a significant impact on the performance of the gate as long as the coupler remains below \SI{3.9}{\giga\hertz}.

Next, we measure the dephasing times of the three hybridized modes as a function of the coupler frequency, see Fig.~\ref{fig:fig_4}(c). At each of the frequencies, we perform a Ramsey experiment and extract the dephasing times from a fit to the model
\begin{equation}
\label{eq:decay_envelope}
    \nu(t) = a(t)\mathrm{e}^{-\frac{t}{2 T_1} -\frac{t}{T_{\phi, 1}} -\left(\frac{t}{T_{\phi, 2}}\right)^2},
\end{equation}
where $a(t)$ is the response of the Ramsey experiment excluding decoherence, $T_1$ is the energy-relaxation time measured earlier, and $T_{\phi, 1}$ and $T_{\phi, 2}$ are the exponential and Gaussian components of the dephasing time. The exponential dephasing time originates from excess wide-band noise such as thermal excitations in the readout resonators \cite{yan2018distinguishing} and is expected to only moderately contribute to the total dephasing. Indeed, we find $T_{\phi, 1}^{\rm Q1} \approx \SI{70}{\micro\second}$ and $T_{\phi, 1}^{\rm Q2} \approx \SI{45}{\micro\second}$ for the coupler frequencies relevant for the CZ gate, see Appendix~\ref{sec:coherence} for the full data. The Gaussian component of the dephasing time is dominated by low-frequency flux noise \cite{bylander2011noise} (see Appendix~\ref{sec:coupler_characterization} for details), which is a function of flux-dispersion of the uncoupled modes and the degree of their hybridisation. We model the dephasing rate of the coupled mode $k$ as
\begin{equation}
    \Gamma_{\phi, 2}^{k} = 1/T_{\phi, 2}^{k} =\sqrt{\sum_{i=\rm{Q_1}, \rm{Q_2}, \rm{c}} \left(p_{i, k}\tilde{\Gamma}_{\phi, 2}^i\right)^2},
\end{equation}
where $p_{i, k}$ is the participation ratio of the uncoupled mode $i$ in the hybridized mode $k$ and $\tilde{\Gamma}_{\phi, 2}^i$ are the dephasing rates of the uncoupled modes, see Fig. \ref{fig:fig_4}(c) for the comparison of the model and the experimental data. Above we have assumed that flux noise affecting the uncoupled modes originates from independent sources and can be approximated as quasi-static noise, resulting in the dephasing rates of the uncoupled modes to sum in square \cite{Ithier2005}. To account for the change in the flux noise sensitivity of the uncoupled coupler mode as its frequency is varied, we model the coupler dephasing rate as $\tilde{\Gamma}_{\phi, 2}^{\rm c} = B|\frac{\partial\omega_{\rm c}}{\partial\Phi_{\rm c}}|$ \cite{bylander2011noise}, where $\omega_{\rm c}/2\pi$ is the frequency of the uncoupled coupler mode, $\Phi_{\rm c}$ is the coupler flux, $B\approx\SI{53}{\Phi_0}$ is the flux noise amplitude extracted from a fit to the coupler dephasing rate when it is decoupled from the qubits, and $\Phi_0$ is the flux quantum. With these assumptions, we observe an excellent match between the experimental data and the model across the measured coupler frequency range, and note that the dephasing times of the qubit modes stay above $\SI{1}{\micro\second}$ even when maximally hybridized with the coupler due to having the coupler flux sweet spot being located near the qubit frequencies.

In order to estimate the coherence limit of an arbitrarily long CZ gate, we calculate effective relaxation times $T_{1, {\rm eff}}$ and effective dephasing times $T_{\phi, 1, {\rm eff}}$ and $T_{\phi, 2, {\rm eff}}$ for both qubits by weighting the measured coherence rates with the time spend on the corresponding continuous adiabatic trajectories [Fig.~\ref{fig:fig_4}(b)] during the CZ gate, see Appendix~\ref{sec:coherence} for additional information. The time spent on each frequency is determined by the Slepian pulse shape, which we parameterize with the interaction time $\tau$. For each $\tau$, we re-calculate the pulse shape using the $ZZ$ interaction strength calibration data in Fig.~\ref{fig:fig_2}(b) to ensure that a conditional phase of $\phi_{11}=\pi$ is accumulated.

The coherence limit can then be calculated as \cite{chu2021, abad2021universal}
\begin{equation}
\label{eq:coherence_limit}
    \epsilon_{\rm CZ}^{\rm limit}(\tau) = \sum_{i=1,2} \frac{2}{5}\left[ \frac{\tau}{T_{1, {\rm eff}}^{{\rm Q}_i} (\tau)} +
    \frac{\tau}{T_{\phi, 1, {\rm eff}}^{{\rm Q}_i} (\tau)} +
    \left(\frac{\tau}{T_{\phi, 2, {\rm eff}}^{{\rm Q}_i}(\tau)}\right)^2 \right],
\end{equation}
where the first term in the square brackets corresponds to the coherence limit of the CZ gate due to the relaxation times of qubits and the second and third terms yield the coherence limits due to exponential and Gaussian components of dephasing, respectively. Figure~\ref{fig:fig_4}(d) depicts the coherence limit for the gate error $\epsilon_{\rm CZ}^{\rm limit}$ for various different gate durations along with the contributions of the individual terms in Eq.~\eqref{eq:coherence_limit}.

For our CZ gate with the interaction time of \SI{22}{\nano\second} and an additional \SI{11}{\nano\second} wait time around it, we obtain $T_{1, {\rm eff}}^{{\rm Q}_1} = \SI{14}{\micro\second}$, $T_{1, {\rm eff}}^{{\rm Q}_2} = \SI{43}{\micro\second}$, $T_{\phi, 1, {\rm eff}}^{{\rm Q}_1} = \SI{67}{\micro\second}$, $T_{\phi, 1, {\rm eff}}^{{\rm Q}_2} = \SI{43}{\micro\second}$, $T_{\phi, 2, {\rm eff}}^{{\rm Q}_1} = \SI{17}{\micro\second}$ and $T_{\phi, 2, {\rm eff}}^{{\rm Q}_2} = \SI{6}{\micro\second}$, yielding a coherence limit of $\epsilon_{\rm CZ}^{\rm limit} = \num{1.7e-3}$. This limit can be compared with the experimentally measured gate error $\epsilon_{\rm CZ} = \num[separate-uncertainty]{1.9(2)e-3}$ which is only slightly higher, indicating that the gate is mostly coherence limited, as was already suggested by the randomized benchmarking experiment with multiple interleaved CZ gates. The most significant contribution to the errors comes from the energy relaxation time $T_{1, {\rm eff}}^{{\rm Q}_1}$. Importantly, the fidelity is only modestly affected by dephasing, mainly due to the possibility to operate the coupler near its flux insensitive sweep spot during the CZ gate. To demonstrate that the coupler coherence times at the two-qubit-gate operation region are high enough not to have a detrimental impact on the gate errors, we re-compute the coherence limit assuming that hybridization with the coupler does not affect the qubit coherence times and use this as a reference, see Fig. \ref{fig:fig_4}(d). Comparing the reference case with the coherence limit $\epsilon_{\rm CZ}^{\rm limit}$, we find a negligible difference, confirming the hypothesis.

Possible reasons for the remaining errors include system fluctuations due to TLSs nearby the qubits, leading to rare uncontrollable jumps in the qubit frequencies, as well as uncompensated flux pulse distortions in the nanosecond range and leakage to non-computational states.

\section{Conclusions}

We have proposed and demonstrated a tunable-coupler design based on a floating transmon coupled to two computational qubits via waveguide extenders. This design allows the qubits to be separated at least by \SI{2}{\milli\meter}, enabling each qubit to have individual readout resonators and Purcell filters for high-fidelity readout, low non-nearest-neighbor coupling, and reduced crosstalk from passing control lines to the qubits in a flip-chip architecture. Although there is a minor trade-off between qubit-qubit distance and the coupling strengths due to the added capacitance of the waveguide extenders to ground, we have shown that the interaction strength is large enough for a qubit-qubit distance beyond \SI{1}{\milli\meter} by demonstrating a fast and high-fidelity CZ gate with \SI[separate-uncertainty = true]{99.81(2)}{\percent} fidelity. We confirmed that the CZ gate fidelity is not limited by the coupler coherence, but mostly determined by the qubit energy-relaxation rate, making state-of-the-art two-qubit-gate fidelities on large quantum processors based on this coupler design an achievable goal.

\section{Acknowledgement}
We acknowledge Matthew Sarsby, Roope Kokkoniemi, Ali Yurtalan Jean-Luc Orgiazzi, Lucas Ortega, Jorge Santos, Jaakko Jussila, Illari Kuronen, Jaakko Salo, Tiina Naaranoja, Otto Koskinen, and Tero Somppi for supporting the conceptualization, construction, and maintenance of the experimental setup, Ferenc D\'osa-R\'acz, Janne M\"antyl\"a, Sinan Inel, and Leon Wubben for additional software support, and Olli-Pentti Saira for valuable discussions. We would additionally like to thank the rest of the IQM team for creating the entire infrastructure, laying the foundation of this work.

A.V. and J.HE. conceptualized the project, F.M., S.J., and A.V. planned and executed the  experiments and analyzed the experimental data, C.O.-K. and A.L., conducted and analyzed the microwave simulations, J.T., M.PAP., A.V., M.T., and A.A. provided theoretical modeling and simulations, C.O.-K., A.L., E.T., J.R\"A., and J.HE. designed and simulated the sample, F.M., J.I., T.H., M.K., J.RO., N.V., M.SE., L.B., O.A., B.T., V.B., J.K., D.J., E.H., C.F.C., M.SA., V.S., M.PAR., F.T., J.L., J.HE., A.V., and S.J. developed the experiment and analysis software, 
W.L. and T.L. designed Josephson Junctions. W.L. developed qubit and airbridge process, fabricated the qubit devices, W.L. and Y.L. benchmarked the room temperature resistance, J.HO bonded and packaged the device, F.M., A.V., S.J., J.T., A.L., C.O.-K., W.L, M.M., and J.HE. drafted or revised the work, and J.HE., M.M., J.HA., and K.Y.T. supervised the work. 
Experimental data are available upon reasonable request to the authors.

The work was partly supported by the European Innovation Council (EIC) under Prometheus (grant No.  959521), Business Finland (grant No. 7547/31/2021), and by the German Federal Ministry of Education and Research (BMBF) under Q-Exa (grant No. 13N16062), QSolid (grant No. 13N16161), and MUNIQC-SC (grant No. 13N16185). M.M. is partly supported by the Academy of Finland through its Centers of Excellence Program (project No. 336810) and by the European Research Council under Advanced Grant ConceptQ (grant No. 101053801). Parts of this work are included in patents applications filed by IQM Finland Oy. This work used resources from the OtaNano Micronova cleanroom.
\bibliography{ref}
\pagebreak
\clearpage

\onecolumngrid
\title{Long-distance transmon coupler with CZ gate fidelity above \SI{99.8}{\percent}}

\maketitle

\onecolumngrid
\clearpage

\setcounter{equation}{0}
\setcounter{figure}{0}
\setcounter{table}{0}
\setcounter{page}{1}
\setcounter{section}{0}

\newcolumntype{C}{>{\centering\arraybackslash} m{1.4cm} }

\renewcommand{\theequation}{S\arabic{equation}}
\renewcommand{\thefigure}{S\arabic{figure}}
\renewcommand{\theHfigure}{S\arabic{figure}}
\renewcommand{\bibnumfmt}[1]{[S#1]}
\newcommand{\note}[1]
  {\begingroup{\color{blue}[NOTE: \textit{#1}]}\endgroup}
\appendix

\section{EFFECTIVE COUPLING MODEL}
\label{sec:effective_coupling}

\begin{figure*}
    \includegraphics[width=0.55\linewidth]{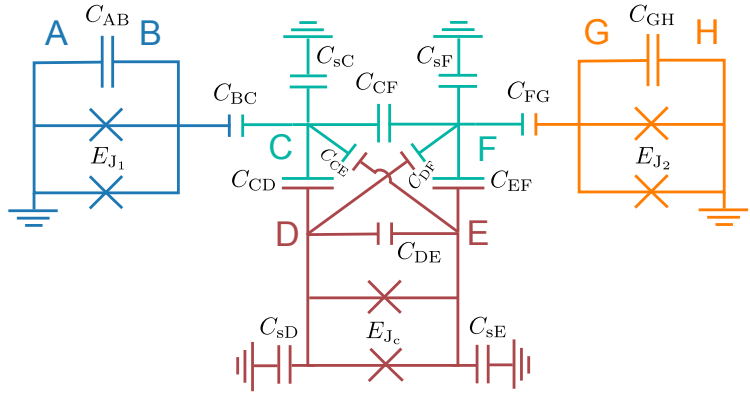}
    \caption{Lumped-element circuit diagram of the floating-coupler scheme. The drawn sizes of the capacitors roughly indicate the relative magnitudes between the designed values of the capacitances. The crosses indicate the Josephson junctions with the Josephson energy $E_{\rm J}$.}\label{fig:schematic_floating_coupler}
\end{figure*}

In this Appendix, we derive an effective Hamiltonian operator for the extended floating-coupler setup. The essential physics can be described with the lumped-element circuit shown in Fig.~\ref{fig:schematic_floating_coupler}. Since the microwave wavelengths considered here are much longer than the lengths of the waveguide extenders, the extenders can be modeled as lumped capacitors for the derivation of the effective Hamiltonian, as shown in Fig.~\ref{fig:schematic_floating_coupler}. We also observe that the shunt capacitances $C_{\rm sC}$, $C_{\rm sD}$, $C_{\rm sE}$, and $C_{\rm sF}$ to the ground, and the coupling capacitances $C_{\rm CE}$ and $C_{\rm DF}$ to opposite coupler islands are parasitic couplings, hence not needed to operate the two-qubit system. However, such capacitances are unavoidably present in our design, and thus, also included in the discussion below.

The derivation of the Hamiltonian follows the conventional circuit-quantization procedure described in the sections below. However, we note that the circuit has six voltage nodes, three of which can be eliminated. We eliminate the two (inactive) purely capacitive nodes C and F using three successive star-mesh transformations. Consequently, we write down the Lagrangian of the circuit and show that the relative and the "center-of-mass" motion of the coupler can be separated. The center-of-mass coordinate is cyclic, and thus can also be eliminated. The resulting Hamiltonian has three degrees of freedom, one for each qubit and the coupler. Finally, we show how the effective $ZZ$ interaction strength can be computed numerically using the obtained effective Hamiltonian operator.

\subsection{Elimination of the inactive nodes}
\begin{figure*}
\includegraphics[width=0.65\linewidth]{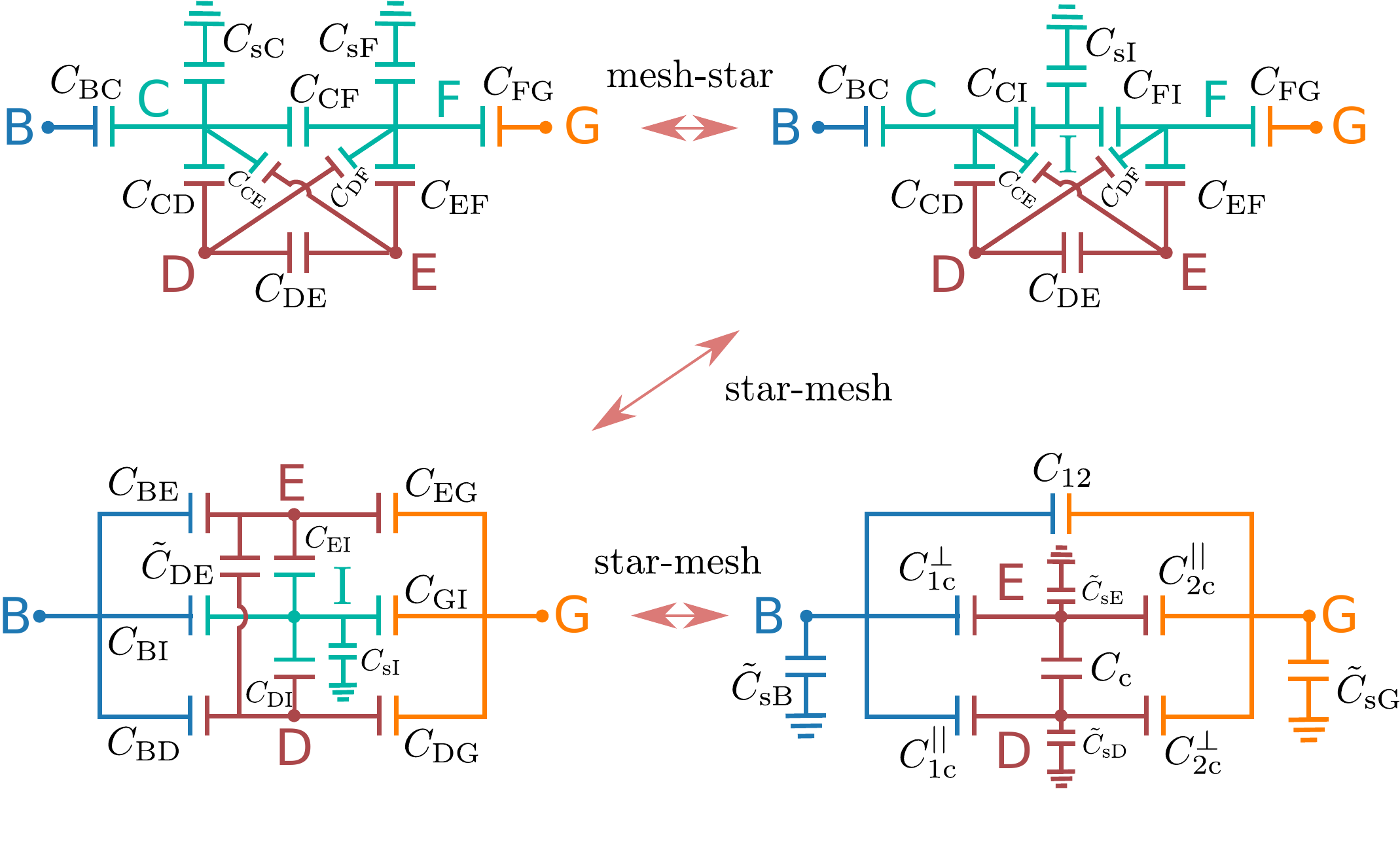}
\caption{Derivation of the effective model using star-mesh transformations. The capacitances are defined in Eqs.~\eqref{eq:caps} and~\eqref{eq:caps2}.
}\label{fig:transformations}
\end{figure*}

Here, we observe that in the circuit shown in Fig.~\ref{fig:schematic_floating_coupler}, the nodes C and F are inactive because they are coupled to other nodes only capacitively. In Fig.~\ref{fig:transformations}, we show that the direct coupling between nodes C and F can be eliminated by first applying the mesh-star ($\Delta$-Y) transformation for the nodes C, F, and ground, defined as
\begin{equation}
    \begin{split}
        C_{\rm CI} &= \frac{C_{\rm CF}C_{\rm sC} + C_{\rm CF}C_{\rm sF} + C_{\rm sC}C_{\rm sF}}{C_{\rm sF}},\\
        C_{\rm FI} &= \frac{C_{\rm CF}C_{\rm sC} + C_{\rm CF}C_{\rm sF} + C_{\rm sC}C_{\rm sF}}{C_{\rm sC}},\\
        C_{\rm sI} &= \frac{C_{\rm CF}C_{\rm sC} + C_{\rm CF}C_{\rm sF} + C_{\rm sC}C_{\rm sF}}{C_{\rm CF}}.
    \end{split}
\end{equation}
Consequently, the coupling is mediated by an effective node I. Then, we apply the star-mesh transformations to the nodes C and F, resulting in the desired elimination of the nodes. Subsequently, we eliminate the node I with an of the star-mesh transformation. The star-mesh transformation is defined in the general case as
\begin{equation}
    C_{\rm XY} = \frac{C_{\rm X}C_{\rm Y}}{C_{||}},
\end{equation}
where $C_{||}=\sum_n C_n$ is the sum of all capacitances $C_n$ coupled to the central node in the star configuration, and $C_{\rm XY}$ is the effective capacitance coupling nodes $X$ and $Y$ in the mesh configuration.

\begin{figure*}
\includegraphics[width=0.55\linewidth]{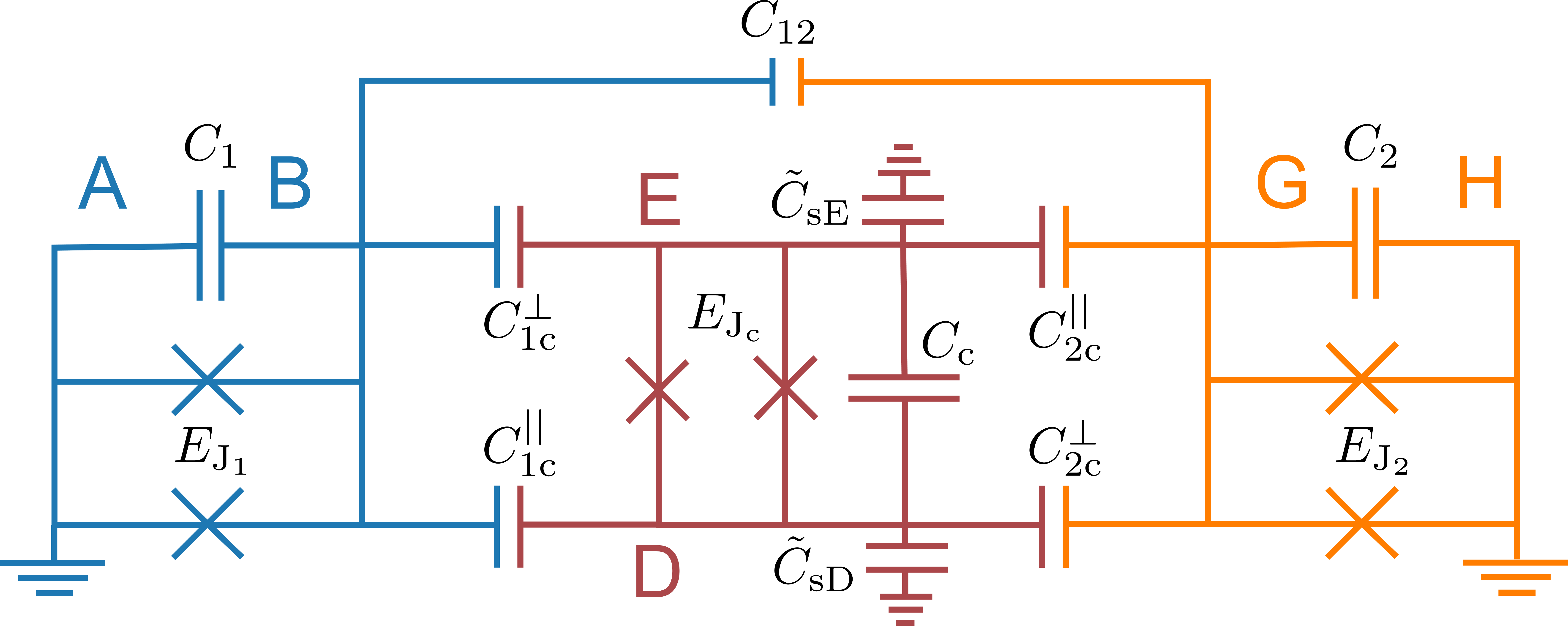}
\caption{Effective-circuit schematic of the floating-coupler setup. The capacitances are defined in Eq.~\eqref{eq:caps}.
}\label{fig:reduced_floating_coupler}
\end{figure*}

Consequently, we obtain the equivalent circuit shown in Fig.~\ref{fig:reduced_floating_coupler}, in which the inactive nodes C and F have been eliminated. The related effective capacitances can be written in terms of the physical capacitances as
\begin{equation}\label{eq:caps}
\begin{split}
C_1 &= C_{\rm AB} + \frac{C_{\rm BI}C_{\rm sI}}{C_{||}^{\rm I}}\\ 
&\approx C_{\rm AB} + C_{\rm sC}\frac{C_{\rm BC}}{C_{||}^{\rm C}}\left[1+\frac{C_{\rm CF}}{C_{\rm ||}^{\rm F}}\frac{C_{\rm FI}}{C_{\rm CI}}+\frac{C_{\rm CF}}{C_{\rm ||}^{\rm C}}\frac{C_{\rm CI}}{C_{\rm FI}}\right],\\
C_2 &= C_{\rm GH} + \frac{C_{\rm GI}C_{\rm sI}}{C_{||}^{\rm I}}\\
&\approx C_{\rm GH} + C_{\rm sF}\frac{C_{\rm FG}}{C_{||}^{\rm F}}\left[1+\frac{C_{\rm CF}}{C_{||}^{\rm F}}\frac{C_{\rm FI}}{C_{\rm CI}}+\frac{C_{\rm CF}}{C_{\rm ||}^{\rm C}}\frac{C_{\rm CI}}{C_{\rm FI}}\right],\\
C_{\rm c} &= C_{\rm DE}+ \frac{C_{\rm CD}C_{\rm CE}}{C_{||}^{\rm C}} + \frac{C_{\rm DF}C_{\rm EF}}{C_{||}^{\rm F}}+ \frac{C_{\rm DI}C_{\rm EI}}{C_{||}^{\rm I}}\\
&\approx C_{\rm DE} + \frac{C_{\rm CD}C_{\rm EF}\tilde{C}_{\rm CF}}{C_{||}^{\rm C}C_{||}^{\rm F}},\\
\tilde{C}_{\rm sD} &= C_{\rm sD} + \frac{C_{\rm DI}C_{\rm sI}}{C_{||}^{\rm I}} \\
&\approx C_{\rm sD} + C_{\rm sC}\frac{C_{\rm CD}}{C_{||}^{\rm C}}\left[1+\frac{C_{\rm CF}}{C_{||}^{\rm F}}\frac{C_{\rm FI}}{C_{\rm CI}}+\frac{C_{\rm CF}}{C_{\rm ||}^{\rm C}}\frac{C_{\rm CI}}{C_{\rm FI}}\right],\\
\tilde{C}_{\rm sE} &= C_{\rm sE} + \frac{C_{\rm EI}C_{\rm sI}}{C_{||}^{\rm I}}\\
& \approx C_{\rm sE} + C_{\rm sF}\frac{C_{\rm EF}}{C_{||}^{\rm F}}\left[1+\frac{C_{\rm CF}}{C_{||}^{\rm F}}\frac{C_{\rm FI}}{C_{\rm CI}}+\frac{C_{\rm CF}}{C_{\rm ||}^{\rm C}}\frac{C_{\rm CI}}{C_{\rm FI}}\right].
\end{split}
\qquad \qquad
\begin{split}
C_{\rm 1c}^{||} &= \frac{C_{\rm BC}C_{\rm CD}}{C_{||}^{\rm C}} + \frac{C_{\rm BI}C_{\rm DI}}{C_{||}^{\rm I}}\\ 
&\approx \frac{C_{\rm BC}C_{\rm CD}}{C_{||}^C}\left[1+ \frac{C_{\rm CI}}{C_{\rm FI}}\frac{\tilde{C}_{\rm CF}}{C_{||}^{\rm C}}\right],\\
C_{\rm 1c}^{\perp} &= \frac{C_{\rm BC}C_{\rm CE}}{C_{||}^{\rm C}}+\frac{C_{\rm BI}C_{\rm EI}}{C_{||}^{\rm I}}\\ 
&\approx \left[C_{\rm CE} + \frac{C_{\rm EF}C_{\rm CF}}{C_{||}^{\rm F}}\right]\frac{C_{\rm BC}}{C_{||}^{\rm C}},\\
C_{\rm 2c}^{||} &= \frac{C_{\rm EF}C_{\rm FG}}{C_{||}^{\rm F}} + \frac{C_{\rm EI}C_{\rm GI}}{C_{||}^{\rm I}}\\
&\approx \frac{C_{\rm EF}C_{\rm FG}}{C_{||}^{\rm F}}\left[1+\frac{C_{\rm FI}}{C_{\rm CI}}\frac{\tilde{C}_{\rm CF}}{C_{||}^{\rm F}}\right],\\
C_{\rm 2c}^{\perp} &= \frac{C_{\rm DF}C_{\rm FG}}{C_{||}^{\rm F}}+\frac{C_{\rm DI}C_{\rm GI}}{C_{||}^{\rm I}} \\
&\approx \left[C_{\rm DF} + \frac{C_{\rm CD}C_{\rm CF}}{C_{||}^{\rm C}}\right]\frac{C_{\rm FG}}{C_{||}^{\rm F}},\\
C_{12} &= \frac{C_{\rm BI}C_{\rm GI}}{C_{||}^{\rm I}} = \frac{C_{\rm BC}\tilde{C}_{\rm CF}C_{\rm FG}}{C_{||}^{\rm C}C_{||}^{\rm F}},
\end{split}
\end{equation}
where the effective capacitances that arise from the mesh-star and star-mesh transformations are defined as
\begin{equation}\label{eq:caps2}
\begin{split}
C_{\rm CI} &= C_{\rm CF} + C_{\rm sC} + \frac{C_{\rm CF}C_{\rm sC}}{C_{\rm sF}},\\
C_{\rm FI} &= C_{\rm CF} + C_{\rm sF} + \frac{C_{\rm CF}C_{\rm sF}}{C_{\rm sC}},\\
C_{\rm sI} &= C_{\rm sC} + C_{\rm sF} + \frac{C_{\rm sC}C_{\rm sF}}{C_{\rm CF}},\\
C_{\rm BD} &= \frac{C_{\rm BC}C_{\rm CD}}{C_{||}^{\rm C}},\\
C_{\rm BE} &= \frac{C_{\rm BC}C_{\rm CE}}{C_{||}^{\rm C}},\\
C_{\rm BI} &= \frac{C_{\rm BC}C_{\rm CI}}{C_{||}^{\rm C}},\\
C_{\rm DI} &= \frac{C_{\rm DF}C_{\rm FI}}{C_{||}^{\rm F}}+ \frac{C_{\rm CD}C_{\rm CI}}{C_{||}^{\rm C}},\\
C_{||}^{\rm C} &= C_{\rm BC} + C_{\rm CD} + C_{\rm CE} + C_{\rm CF} + C_{\rm sC} + \frac{C_{\rm CF} C_{\rm sC}}{C_{\rm sF}},\\
\end{split}
\qquad \qquad
\begin{split}
\tilde{C}_{\rm DE} &= C_{\rm DE} + \frac{C_{\rm CD}C_{\rm CE}}{C_{||}^{\rm C}} + \frac{C_{\rm DF}C_{\rm EF}}{C_{||}^{\rm F}},\\
C_{\rm DG} &= \frac{C_{\rm DF}C_{\rm FG}}{C_{||}^{\rm F}},\\
C_{\rm EG} &= \frac{C_{\rm EF}C_{\rm FG}}{C_{||}^{\rm F}},\\
C_{\rm EI} &= \frac{C_{\rm CI}C_{\rm CE}}{C_{||}^{\rm C}} + \frac{C_{\rm EF}C_{\rm FI}}{C_{||}^{\rm F}},\\
C_{\rm GI} &= \frac{C_{\rm FG}C_{\rm FI}}{C_{||}^{\rm F}},\\
C_{||}^{\rm F} &= C_{\rm DF} + C_{\rm EF} + C_{\rm FG} + C_{\rm CF} + C_{\rm sF}  + \frac{C_{\rm CF}C_{\rm sF}}{C_{\rm sC}},\\
C_{||}^{\rm I} &= C_{\rm BI} + C_{\rm DI} + C_{\rm EI} + C_{\rm GI} + C_{\rm sI},\\
\tilde{C}_{\rm CF} &= \frac{C_{\rm CI}C_{\rm FI}}{C_{||}^{\rm I}} = \frac{C_{\rm CF}}{1-C_{\rm CF}\left[\frac{C_{\rm CI}}{C_{\rm FI}C_{||}^{\rm C}}+ \frac{C_{\rm FI}}{C_{\rm CI}C_{||}^{\rm F}}\right]}.
\end{split}
\end{equation}
Above, the approximations hold if $C_{\rm CE},C_{\rm CI} \ll C_{||}^{\rm C}$ and $C_{\rm DF},C_{\rm FI} \ll C_{||}^{\rm F}$.

An alternative approach to eliminate the inactive nodes C and F in Fig.~\ref{fig:schematic_floating_coupler} is to calculate the admittance matrix $Y$ of a four-port network, which includes the nodes B, D, E and G. We can model the network by connecting each node to every other node with a lumped element with a certain admittance $y_{ij}$, where $i, j \in \{\rm{B, D, E, G}\}$. Each admittance $y_{ij}$ can be calculated at a certain frequency $\omega$ from the initial capacitances shown in Fig.~\ref{fig:schematic_floating_coupler}. The admittance matrix can then be calculated as
\begin{equation}
    Y_{ij} = 
    \begin{cases}
    y_{i} + \sum_{k=\rm{B, D, E, G} \atop k \neq i} {y_{ik}}, & \mbox{if} \quad i = j \\
    -y_{ij},  & \mbox{if} \quad i \neq j \quad .
    \end{cases}
\end{equation}
Assuming having only capacitive elements in the network, we can then calculate the capacitance matrix as $C = \rm{Im} (Y(\omega)) / \omega$ at a certain angular frequency $\omega$, where $C_{ii}$ is the capacitance from node $i$ to ground and $C_{ij}= C_{ji}$ is the capacitance from node $i$ to node $j$.

\subsection{Elimination of the center of mass}

Here, we write the Lagrangian of the system as $L=T-V$ in which
\begin{equation}\label{eq:2island_kinetic}
\begin{split}
T =& \frac12 C_{1}\dot{\phi}_{\rm B}^2 + 
\frac12 C_{\rm 1c}^{||}(\dot{\phi}_{\rm B}-\dot{\phi}_{\rm D})^2 +
\frac12 C_{\rm 1c}^{\perp}(\dot{\phi}_{\rm B}-\dot{\phi}_{\rm E})^2+
\frac12 C_{\rm 12}(\dot{\phi}_{\rm B}-\dot{\phi}_{\rm G})^2\\ 
&+ \frac12 C_{2}\dot{\phi}_{\rm G}^2 + \frac12 C_{\rm 2c}^{||}(\dot{\phi}_{\rm E}-\dot{\phi}_{\rm G})^2  + \frac12 C_{\rm 2c}^{\perp}(\dot{\phi}_{\rm D}-\dot{\phi}_{\rm G})^2 +
\frac12 C_{\rm c}(\dot{\phi}_{\rm D}-\dot{\phi}_{\rm E})^2 +
\frac12 \tilde{C}_{\rm sD}\dot{\phi}_{\rm D}^2 +
\frac12 \tilde{C}_{\rm sE}\dot{\phi}_{\rm E}^2,
\end{split}
\end{equation}
and
\begin{equation}\label{eq:2island_potential}
V = E_{\rm J_1}\left(1-\cos\left[\frac{2\pi}{\Phi_0}\phi_{\rm B}\right]\right) +
E_{\rm J_2}\left(1-\cos\left[\frac{2\pi}{\Phi_0}\phi_{\rm G}\right]\right) +
E_{\rm J_{\rm c}}\left(1-\cos\left[\frac{2\pi}{\Phi_0}(\phi_{\rm D}-\phi_{\rm E})\right]\right),
\end{equation}
where $\Phi_0 = h/(2e)$ is the flux quantum and $\phi_i$ is the node flux related to the node $i\in\{\rm B,D,E,G\}$. We emphasize that both qubits and the coupler have a charging energy $E_{\rm C}\ll E_{\rm J}$, and thus operated in the transmon regime. Consequently, we have neglected the offset charges from the above expression for kinetic energy, as their detrimental effect to quantum coherence is exponentially suppressed in this regime of parameters. 

We observe that the potential energy depends only on the relative motion $\phi_{\rm c} \equiv \phi_{\rm D}-\phi_{\rm E}$. This suggests that there exists a constant of motion analogous to the center-of-mass in mechanical systems. Therefore, we make a linear transformation $(\phi_{\rm D}, \phi_{\rm E}) \rightarrow (\phi_{\rm c}, \theta_{\rm c})$ to flux coordinates of relative motion, defined as 
\begin{equation}
    \left(\begin{array}{cc}
         \phi_{\rm c}  \\
         \theta_{\rm c} 
    \end{array}\right) =
    \left(\begin{array}{cc}
         1 & -1 \\
         a & b
    \end{array}\right)
    \left(\begin{array}{cc}
         \phi_{\rm D}  \\
         \phi_{\rm E} 
    \end{array}\right).
\end{equation}
Below, we require that the determinant of the transformation matrix is equal to $a+b = 1$, such that there is no scaling involved. Moreover, we define the transformation such that the fluxes $\phi_{\rm c}$ and $\theta_{\rm c}$ are decoupled.
Consequently, we obtain that
\begin{equation}
\theta_{\rm c} \equiv \frac{(C_{\rm 1c}^{||}+C_{\rm 2c}^{\perp}+\tilde{C}_{\rm sD})\phi_{\rm D} + (C_{\rm 1c}^{\perp} + C_{\rm 2c}^{||} + \tilde{C}_{\rm sE})\phi_{\rm E}}{C_{\rm 1c}^{||}+C_{\rm 1c}^{\perp}+C_{\rm 2c}^{||}+C_{\rm 2c}^{\perp}+\tilde{C}_{\rm sD}+\tilde{C}_{\rm sE}},    
\end{equation}
which is a cyclic coordinate. We observe that the flux $\theta_{\rm c}$ can be interpreted as the "center of capacitance" of the two islands D and E, in which the total (decoupled) island capacitances are given as sums of the capacitances coupled to the island, i.e. $C_{\rm 1c}^{||}+C_{\rm 2c}^{\perp}+\tilde{C}_{\rm sD}$ and $C_{\rm 1c}^{\perp}+C_{\rm 2c}^{||}+\tilde{C}_{\rm sE}$, respectively.

Denoting $\phi_1 = \phi_{\rm B}$ and $\phi_2=\phi_{\rm G}$, the kinetic energy defined in Eq.~\eqref{eq:2island_kinetic} can be written after the transformation as $T=\frac12 \dot{\Phi}^{\rm T} C \dot{\Phi}$, where $\Phi^{\rm T}=(\phi_1,\phi_{\rm c},\theta_{\rm c},\phi_2)$, and the corresponding capacitance matrix is given as
\begin{equation}\label{eq:centermassCap}
C=\left(\begin{array}{cccc}
C_{\sigma_1} & -C_{\rm 1c} & -C_{\rm 1c}^{||}-C_{\rm 1c}^{\perp} & -C_{12}\\
-C_{\rm 1c} & C_{\sigma_{\rm c}} & 0 & C_{\rm 2c}\\
-C_{\rm 1c}^{||}-C_{\rm 1c}^{\perp} & 0 & C_{\sigma_{\theta}} & -C_{\rm 2c}^{||}-C_{\rm 2c}^{\perp}\\
-C_{12} & C_{\rm 2c} & -C_{\rm 2c}^{||}-C_{\rm 2c}^{\perp} & C_{\sigma_2}
\end{array}\right), 
\end{equation}
where 
\begin{equation}
\begin{split}
C_{\sigma_1} &= C_{1} + C_{\rm 1c}^{||} + C_{\rm 1c}^{\perp}+ C_{12},\\
C_{\sigma_{\rm c}} &= C_{\rm c} + \left[\frac{1}{C_{\rm 1c}^{||} + C_{\rm 2c}^{\perp} + \tilde{C}_{\rm sD}} + \frac{1}{C_{\rm 2c}^{||} + C_{\rm 1c}^{\perp} + \tilde{C}_{\rm sE}}\right]^{-1} = C_{\rm c} + \gamma_1\gamma_2 C_{\sigma_{\theta}},\\
C_{\sigma_2} &= C_{2} + C_{\rm 2c}^{||} + C_{\rm 2c}^{\perp} + C_{12},\\
C_{\sigma_{\theta}} &= C_{\rm 1c}^{||} + C_{\rm 1c}^{\perp} + C_{\rm 2c}^{||} + C_{\rm 2c}^{\perp} + \tilde{C}_{\rm sD} + \tilde{C}_{\rm sE},\\
C_{\rm 1c} &= \gamma_1C_{\rm 1c}^{||} - \gamma_2C_{\rm 1c}^{\perp},\\
C_{\rm 2c} &= \gamma_2C_{\rm 2c}^{||} - \gamma_1C_{\rm 2c}^{\perp},
\end{split}
\end{equation}
and $\gamma_{1} = (C_{\rm 2c}^{||} + C_{\rm 1c}^{\perp}+\tilde{C}_{\rm sE})/C_{\sigma_{\theta}}$, $\gamma_{2} = (C_{\rm 1c}^{||} + C_{\rm 2c}^{\perp} + \tilde{C}_{\rm sD})/C_{\sigma_{\theta}}$. We point out that the second term in the definition of $C_{\sigma_{\rm c}}$ is equivalent to the reduced mass of the mechanical two-body problem. We observe that after the transformation, the relative and center-of-mass motion are indeed decoupled. However, there is a coupling between the center-of-mass and the qubit fluxes. The potential energy does not depend on the center-of-mass flux which is, thus, a cyclic coordinate. According to the Euler--Lagrange equation, the corresponding conjugate momentum,
\begin{equation}
q_{\theta} = \frac{\partial L}{\partial \dot{\theta}_{\rm c}} = -(C_{\rm 1c}^{||} + C_{\rm 1c}^{\perp})\dot{\phi}_{\rm B} - (C_{\rm 2c}^{||}+C_{\rm 2c}^{\perp})\dot{\phi}_{\rm G} + C_{\sigma_{\theta}}\dot{\theta}_{\rm c}, 
\end{equation}
is a constant of motion, i.e. $q_{\theta} = $ constant. Consequently, we obtain
\begin{equation}
\dot{\theta}_{\rm c} = \frac{1}{C_{\sigma_{\theta}}}\left[q_{\theta} + (C_{\rm 1c}^{||} + C_{\rm 1c}^{\perp})\dot{\phi}_{\rm B} + (C_{\rm 2c}^{||}+C_{\rm 2c}^{\perp})\dot{\phi}_{\rm G}\right].
\end{equation}
Substituting this back to the Lagrangian, and neglecting the constant term proportional to $q_{\theta}$, we obtain the effective capacitance matrix
\begin{equation}\label{eq:effCapMat}
C'=\left(\begin{array}{ccc}
C_{\Sigma_1}& -C_{\rm 1c} & - C_{12}^*\\
-C_{\rm 1c} & C_{\Sigma_{\rm c}} & C_{\rm 2c}\\
-C_{12}^*& C_{\rm 2c} & C_{\Sigma_2}
\end{array}\right), 
\end{equation}
where
\begin{equation}
\begin{split}
C_{\Sigma_1} &= C_{\sigma_1} - \frac{(C_{\rm 1c}^{||}+C_{\rm 1c}^{\perp})^2}{C_{\sigma_{\theta}}},\\
C_{12}^* &= C_{12} + \frac{(C_{\rm 1c}^{||}+C_{\rm 1c}^{\perp})(C_{\rm 2c}^{||}+C_{\rm 2c}^{\perp})}{C_{\sigma_{\theta}}},\\
C_{\Sigma_2} &= C_{\sigma_2} - \frac{(C_{\rm 2c}^{||}+C_{\rm 2c}^{\perp})^2}{C_{\sigma_{\theta}}}.
\end{split}
\end{equation}
We note that the effective coupling capacitance $C_{12}^*$ between the qubits consists of the direct capacitance $C_{12}$ and that mediated by the coupler structure.
The relevant coordinate flux vector is defined as $\Phi' = (\phi_{1},\phi_{\rm c},\phi_{2})$, in which $\phi_1 = \phi_{\rm B}$, $\phi_2 = \phi_{\rm G}$, and $\phi_{\rm c} = \phi_{\rm D}-\phi_{\rm E}$ is the coordinate for the relative motion between the nodes D and E. Similar discussions on mediated interactions in a chain of floating transmons have been presented recently in Refs.~\onlinecite{yanay2022mediated, spurious_crosstalk_2022}.

\subsection{Effective model without cross-island coupling}

\begin{figure*}
\includegraphics[width=0.55\linewidth]{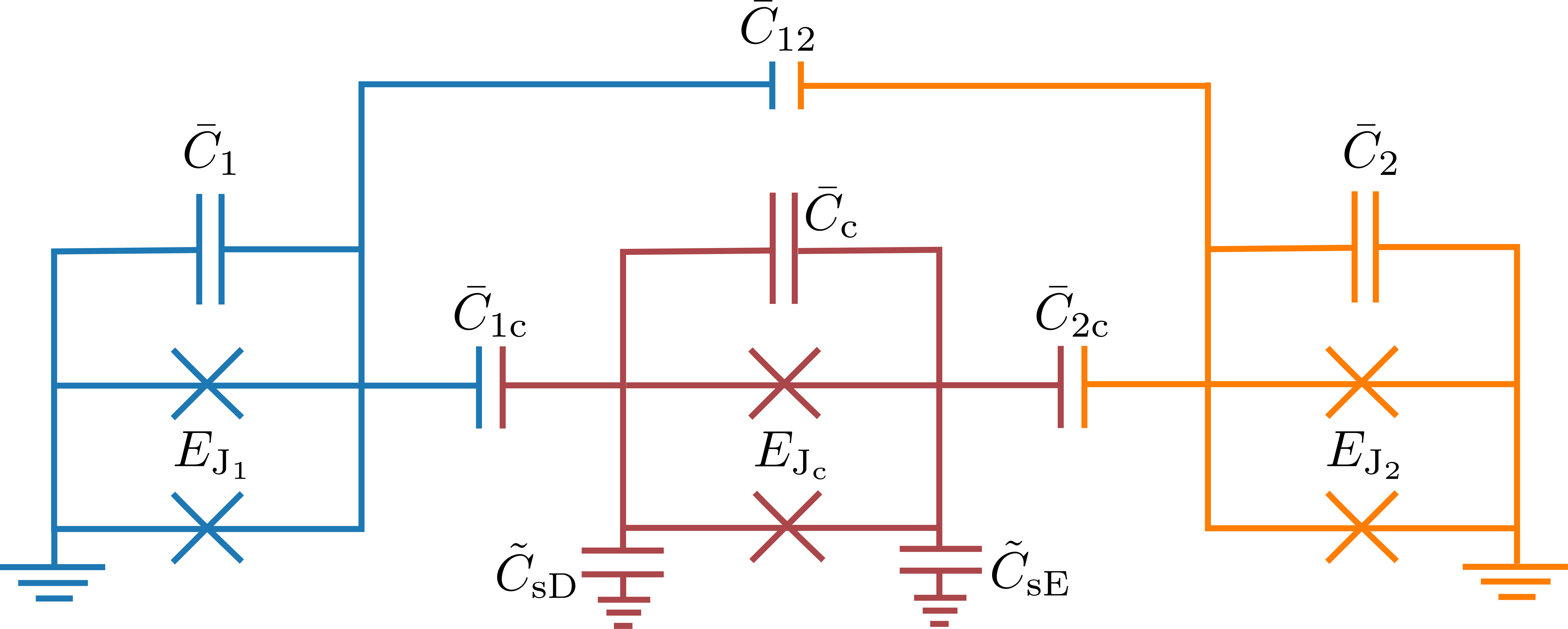}
\caption{Schematic of the reduced effective circuit of the floating-coupler setup. The capacitances are defined in Eq.~\eqref{eq:red_eff_cap}.
}\label{fig:reduced_effective_coupler}
\end{figure*}

Let us consider the effective capacitance matrix in Eq.~\eqref{eq:effCapMat}. We observe that the elements of the matrix depend on cross-island coupling capacitances $C_{\rm 1c}^{\perp}$ and $C_{\rm 2c}^{\perp}$. Here, we show that an equivalent effective capacitance matrix can be realized with a circuit in which $C_{\rm 1c}^{\perp}=C_{\rm 2c}^{\perp}=0$, provided that the other capacitances are adjusted such that the coupling capacitances in Eq.~\eqref{eq:effCapMat} remain unaltered. We find that this can be achieved by making a transformation $C\rightarrow \bar{C}$ to Eq.~\eqref{eq:centermassCap}, in which the elements of $\bar{C}$ are defined as
\begin{equation}\label{eq:red_eff_cap}
\begin{split}
\bar{C}_{\rm 1c}^{\perp} &= \bar{C}_{\rm 2c}^{\perp} = 0,\\
\bar{C}_{\rm 1c}^{||} &= C_{\rm 1c}^{||} - C_{\rm 1c}^{\perp}\frac{C_{\rm 1c}^{||} + C_{\rm 2c}^{\perp}+\tilde{C}_{\rm sD}}{C_{\rm 2c}^{||}+C_{\rm 1c}^{\perp}+\tilde{C}_{\rm sE}},\\
\bar{C}_{\rm 2c}^{||} &=C_{\rm 2c}^{||} - C_{\rm 2c}^{\perp}\frac{C_{\rm 2c}^{||} + C_{\rm 1c}^{\perp}+\tilde{C}_{\rm sE}}{C_{\rm 1c}^{||}+C_{\rm 2c}^{\perp}+\tilde{C}_{\rm sD}},\\
\bar{C}_{\rm 12} &= C_{12}^* - \frac{\bar{C}_{\rm 1c}^{||}\tilde{C}_{\rm 2c}^{||}}{\bar{C}_{\sigma_{\theta}}},\\
\bar{C}_{\sigma_{\theta}} &= \bar{C}_{\rm 1c}+\bar{C}_{\rm 2c} + \tilde{C}_{\rm sD} + \tilde{C}_{\rm sE},\\
\bar{C}_{\sigma_1} &= C_{\Sigma_1} + \frac{\bar{C}_{\rm 1c}^{||2}}{\bar{C}_{\Sigma_{\theta}}},\\
\bar{C}_{\sigma_2} &= C_{\Sigma_2} + \frac{\bar{C}_{\rm 2c}^{||2}}{\bar{C}_{\Sigma_{\theta}}},\\
\bar{C}_{\rm c} &= C_{\Sigma_{\rm c}} - \left[\frac{1}{\bar{C}_{\rm 1c} + \tilde{C}_{\rm sD}}+\frac{1}{\bar{C}_{\rm 2c}+\tilde{C}_{\rm sE}}\right]^{-1}.
\end{split}
\end{equation}
We note here that the transformed capacitance matrix $\bar{C}$ above is equivalent to that shown in Eq.~\eqref{eq:centermassCap} in the sense that it results into formally identical effective capacitance matrix $C'$ given in Eq.~\eqref{eq:effCapMat}, and thus, an identical effective Hamiltonian derived in the following section. Therefore, despite the actual values of the circuit capacitances, one can always describe the system with an effective circuit in which $C_{\rm 1c}^{\perp} = C_{\rm 2c}^{\perp} = 0$. This equivalent circuit diagram of the device is shown in Fig.~\ref{fig:reduced_effective_coupler}, and also in Fig.~\ref{fig:fig_1}(a) of the main text. Note that in the main text $C_{12} \circumeq \bar{C}_{12}$, $C_{\rm 1c} \circumeq \bar{C}_{\rm 1c}$ and $C_{\rm 2c} \circumeq \bar{C}_{\rm 2c}$.

The capacitance values of the capacitances shown in Fig.~\ref{fig:reduced_effective_coupler} are listed in Table~\ref{tab:capacitances}. Comparing the capacitance of $\rm{Q}_1$ to ground, $\bar{C}_1 = \SI{82}{\femto\farad}$, to the combined coupling from $\rm{Q}_1$ to the coupler and $\rm{Q}_2$, $\bar{C}^{||}_{1c} + \bar{C}_{12} = \SI{7.25}{\femto\farad}$, we show that coupling four waveguide extenders to $\rm{Q}_1$ is achievable, since $(\bar{C}^{||}_{1c} + \bar{C}_{12})/\bar{C}_1 \ll 1/4$.

\subsection{Hamiltonian of the setup}

The Hamiltonian of the system is obtained with the Legendre transformation and can be written as
\begin{equation}
H = \frac12 \mathbf{q}C'^{-1}\mathbf{q}^{\rm T} + V,
\end{equation}
where $\mathbf{q} = (q_{1},q_{\rm c},q_{2})$. Here, the canonical charges are defined as
\begin{equation}
q_{\lambda} = \frac{\partial L}{\partial \dot{\phi}_{\lambda}} =\frac{\partial T}{\partial \dot{\phi}_{\lambda}}.
\end{equation}
After inverting the capacitance matrix, we obtain the Hamiltonian as
\begin{equation}
\begin{split}\label{eq:Ham}
H =& \frac{1}{2|C'|}\left[A_{\rm 11}q_{\rm 1}^2 + A_{\rm cc}q_{\rm c}^2 + A_{\rm 22} q_{\rm 2}^2 + 2A_{\rm 1c}q_{\rm 1}q_{\rm c} + 2A_{\rm 12}q_{\rm 1}q_{\rm 2} + 2A_{\rm 2c}q_{\rm 2}q_{\rm c} \right] + V.
\end{split}
\end{equation}

Above, we have defined the inverse capacitance matrix as
\begin{equation}\label{eq:capmat}
C'^{-1} \equiv \frac{1}{|C'|}\left(\begin{array}{ccc}
A_{\rm 11} & A_{\rm 1c} & A_{\rm 12} \\
A_{\rm 1c} & A_{\rm cc} &  A_{\rm 2c} \\
A_{\rm 12} &  A_{\rm 2c} & A_{\rm 22} \\
\end{array}\right),
\end{equation}
where $|\cdot|$ indicates the matrix determinant. In the following, we denote $C_{\Sigma_{ij}}^* \equiv |C'|/A_{ij}$ and $C_{\sigma_{ii}} \equiv C_{\sigma_{i}}$ for $i,j\in \{1,\textrm{c},2\}$.

Here, we provide the expressions for the matrix elements of the inverse capacitance matrix in Eq.~\eqref{eq:capmat} as
\begin{equation}
\begin{split}
A_{11} = & C_{\Sigma_2} C_{\Sigma_{\rm c}} - C_{\rm 2c}^2
\approx  C_{\Sigma_2}C_{\Sigma_{\rm c}},\\
A_{\rm cc} = &  C_{\Sigma_1}C_{\Sigma_2} - (C_{12}^*)^2 \approx C_{\Sigma_1} C_{\Sigma_2},\\
A_{22} = & C_{\Sigma_1}C_{\Sigma_{\rm c}} - C_{\rm 1c}^2 \approx C_{\Sigma_1}C_{\Sigma_{\rm c}},\\
\\
A_{\rm 1c} = & C_{\rm 1c}C_{\Sigma_2}  - C_{\rm 2c}C_{12}^* \approx C_{\rm 1c}C_{\Sigma_2}  \\
A_{\rm 2c} = & - C_{\rm 2c}C_{\Sigma_1} + C_{\rm 1c} C_{12}^* \approx -C_{\rm 2c}C_{\Sigma_1},\\
A_{\rm 12} = & C_{12}^*C_{\Sigma_{\rm c}}  - C_{\rm 1c}C_{\rm 2c},
\end{split}
\end{equation}
and the determinant of the capacitance matrix is given by
\begin{equation}
\begin{split}
|C'| = C_{\Sigma_1}C_{\Sigma_2}C_{\Sigma_{\rm c}} - C_{\rm 1c}^2 C_{\Sigma_2} - C_{\rm 2c}^2C_{\Sigma_1} - (C_{12}^*)^2 C_{\Sigma_{\rm c}} + 2C_{\rm 1c}C_{\rm 2c}C_{12}^* \approx C_{\Sigma_1}C_{\Sigma_2}C_{\Sigma_{\rm c}}.
\end{split}
\end{equation}
For the approximations above, we have assumed that $C_{12}\ll C_{\rm 1c}, C_{\rm 2c} \ll C_{\Sigma_1},C_{\Sigma_2},C_{\Sigma_{\rm c}}$.

We define the effective capacitances
\begin{equation}
\begin{split}
C_{\rm \Sigma_1}^* &= \frac{|C'|}{A_{11}} 
\approx C_{\Sigma_1},\\
C_{\rm \Sigma_2}^* &= \frac{|C'|}{A_{22}} 
\approx C_{\Sigma_2},\\
C_{\rm \Sigma_{\rm c}}^* &= \frac{|C'|}{A_{\rm cc}} 
\approx C_{\Sigma_{\rm c}},\\
C_{\Sigma_{\rm 1c}}^* &= \frac{|C'|}{A_{\rm 1c}} 
\approx \frac{C_{\Sigma_1}C_{\Sigma_{\rm c}}}{C_{\rm 1c}},\\
C_{\Sigma_{\rm 2c}}^* &= \frac{|C'|}{A_{\rm 2c}}  
\approx -\frac{C_{\Sigma_2}C_{\Sigma_{\rm c}}}{C_{\rm 2c}},\\
C_{\Sigma_{12}}^* &= \frac{|C'|}{A_{\rm 12}} 
\approx \frac{C_{\Sigma_1}C_{\Sigma_2}}{C_{12}^*(1-\eta)},\end{split}
\end{equation}
where 
\begin{equation}
\eta=  \frac{C_{\rm 1c}C_{\rm 2c}}{C_{12}^*C_{\Sigma_{\rm c}}}.
\end{equation} 

Consequently, the classical Hamiltonian of the system can be expressed as
\begin{equation}
\begin{split}H =& \frac{q_{1}^2}{2C_{\Sigma_1}^*} + \frac{q_{2}^2}{2C_{\Sigma_2}^*} + \frac{q_{\rm c}^2}{2C_{\Sigma_{\rm c}}^*} 
+ \frac{q_{1}q_{\rm c}}{C_{\Sigma_{\rm 1c}}^*} + \frac{q_{\rm 2}q_{\rm c}}{C_{\Sigma_{\rm 2c}}^*} + \frac{q_{\rm 1}q_{\rm 2}}{C_{\Sigma_{12}}^*} \\
&+ E_{\rm J_1}\left(1-\cos\left[\frac{2\pi}{\Phi_0}\phi_{1}\right]\right) +
E_{\rm J_2}\left(1-\cos\left[\frac{2\pi}{\Phi_0}\phi_{2}\right]\right) +
E_{\rm J_{\rm c}}\left(1-\cos\left[\frac{2\pi}{\Phi_0}\phi_{\rm c}\right]\right),
\end{split}
\end{equation}
We follow the canonical quantization procedure and replace the flux coordinates $(\phi_1,\phi_{\rm c},\phi_2)$ and the corresponding canonically conjugated charges $(q_1,q_{\rm c},q_2)$ with operators as $\phi_{\lambda} \rightarrow \hat \phi_{\lambda}$ and $q_{\lambda} \rightarrow \hat q_{\lambda}$ where $\lambda\in [1,\textrm{c},2]$. The operators obey the commutation relation $[\hat \phi_{\lambda},\hat q_{\lambda}] = \textrm{i}\hbar$. Here, we define the charging energies as $E_{\rm C_{\lambda}} = e^2/(2C_{\Sigma_{\lambda}}^*)$. In the limit $E_{\rm C_{\lambda}}\ll E_{\rm J_{\lambda}}$, the setup is equivalent to three bilinearly coupled transmons. Since the transmon is a weakly anharmonic oscillator, it is beneficial to express the flux and charge operators in terms of the annihilation ($\hat b_{\lambda}$) and creation ($\hat b_{\lambda}^{\dag}$) operators of harmonic oscillators corresponding to the linearized potential of the circuit, defined as
\begin{equation}
\begin{split}
\hat \phi_{\lambda}  =& \frac{\Phi_0}{2\pi}\left(\frac{8E_{\rm C_{\lambda}}}{E_{\rm J_{\lambda}}}\right)^{1/4}\frac{1}{\sqrt{2}}(\hat b^{\dag}_{\lambda}+\hat b_{\lambda}),\\
\hat q_{\lambda} =&  \textrm{i}2e\left(\frac{E_{\rm J_{\lambda}}}{8E_{\rm C_{\lambda}}}\right)^{1/4}\frac{1}{\sqrt{2}}(\hat b^{\dag}_{\lambda}-\hat b_{\lambda}).
\end{split}
\end{equation}
Consequently, assuming that $E_{\rm J_{\lambda}} \gg E_{\rm C_{\lambda}}$, the Hamiltonian operator of the system can be written as
\begin{equation}\label{eq:effHam}
\hat H = \hbar \sum_{\lambda} \left( \omega_{\lambda}\hat b_{\lambda}^{\dag}\hat b_{\lambda} +\frac{\alpha_{\lambda}}{2}\hat b_{\lambda}^{\dag}\hat b_{\lambda}^{\dag}\hat b_{\lambda}\hat b_{\lambda}\right) - \hbar g_{\rm 1c}(\hat b_1^{\dag}-\hat b_1)(\hat b_{\rm c}^{\dag} - b_{\rm c}) + \hbar g_{\rm 2c}(\hat b_2^{\dag}-\hat b_2)(\hat b_{\rm c}^{\dag}-\hat b_{\rm c}) - \hbar g_{12}(\hat b_1^{\dag}-\hat b_1)(\hat b_2^{\dag}-\hat b_2)
\end{equation}
where we have defined the qubit angular frequencies, anharmonicities, and the relevant coupling rates as
\begin{equation}
\label{eq:coupling_strengths}
\begin{split}
\hbar\omega_{\lambda} =& \sqrt{8E_{\rm J_{\lambda}}E_{C_{\lambda}}} - E_{C_{\lambda}}, \\
\hbar \alpha_{\lambda} =& - E_{\rm C_{\lambda}},\\
\hbar g_{\rm 1c} =& \frac{e^2}{\sqrt{2}C_{\Sigma_{\rm 1c}}^*}
\left(\frac{E_{\rm J_1}E_{\rm J_{\rm c}}}{E_{\rm C_{1}}E_{\rm C_{\rm c}}}\right)^{1/4} \approx \frac{\hbar}{2} \frac{C_{\rm 1c}}{\sqrt{C_{\Sigma_1}C_{\Sigma_{\rm c}}}}\sqrt{\omega_1\omega_{\rm c}},\\
\hbar g_{\rm 2c} =& -\frac{e^2}{\sqrt{2}C_{\Sigma_{\rm 2c}}^*}
\left(\frac{E_{\rm J_2}E_{\rm J_{\rm c}}}{E_{\rm C_{2}}E_{\rm C_{\rm c}}}\right)^{1/4}\approx \frac{\hbar}{2} \frac{C_{\rm 2c}}{\sqrt{C_{\Sigma_2}C_{\Sigma_{\rm c}}}}\sqrt{\omega_2\omega_{\rm c}},\\
\hbar g_{12} =& \frac{e^2}{\sqrt{2}C_{\Sigma_{12}}^*}
\left(\frac{E_{\rm J_1}E_{\rm J_{2}}}{E_{\rm C_{1}}E_{\rm C_{2}}}\right)^{1/4}\approx \frac{\hbar}{2} \frac{C_{12}^*(1-\eta)}{\sqrt{C_{\Sigma_1}C_{\Sigma_{\rm 2}}}}\sqrt{\omega_1\omega_{2}},
\end{split}
\end{equation}
respectively. Above, we have again assumed that $C_{12}^*\ll C_{\rm 1c}, C_{\rm 2c} \ll C_{\Sigma_1},C_{\Sigma_2},C_{\Sigma_{\rm c}}$ and also neglected the anharmonicities in the latter equalities. For later use, we define the transmon-frequency independent ratio between the coupling rates as
\begin{equation}
\xi = \frac{2g_{\rm 1c}g_{\rm 2c}}{g_{12}\omega_{\rm c}} = \frac{\eta}{1-\eta}.
\end{equation}
We emphasize that the Hamiltonian in Eq.~\eqref{eq:effHam} is similar to that for the conventional tunable-coupler design derived in Ref.~\cite{yan2018tunable}. However, the sign of the coupling term between $\rm{Q}_2$ and the coupler is different, and consequently, the idling frequency of the coupler is located below the qubit frequencies, as we show below.

We have verified the above analytical results independently using a quasi-lumped-element model based on the full circuit schematic shown in Fig.~\ref{fig:schematic_floating_coupler}. The coupler capacitance matrix has been solved using FEM simulations and the full network model has been synthesized from those components. This way we were able to reproduce the theoretical result.

\subsection{Idling frequency of the coupler}

Similar to the conventional design in Ref.~\cite{yan2018tunable}, the Hamiltonian in Eq.~\eqref{eq:effHam} can be approximately diagonalized up to the second order in the qubit-coupler couplings $g_{\rm 1c}$ and $g_{\rm 2c}$ using the Schrieffer--Wolff transformation. Consequently, the effective transverse coupling strength between the single-excitation qubit states can be expressed as
\begin{equation}
\label{eq:gtildeeq}
\tilde{g} 
= g_{12} - \frac{g_{\rm 1c}g_{\rm 2c}}{2}\left(\frac{1}{\Delta_{\rm 1c}}+\frac{1}{\Delta_{\rm 2c}}-\frac{1}{\Sigma_{\rm 1c}}-\frac{1}{\Sigma_{\rm 2c}}\right),
\end{equation} 
where $\Delta_{i\textrm{c}} = \omega_{i}-\omega_{\rm c}$ and $\Sigma_{i\textrm{c}} = \omega_{i} + \omega_{\rm c}$ with $i \in \{1,2\}$. We note that the effective coupling strength $\tilde{g}$ between the coupler-dressed qubits consists of contributions arising from a direct capacitive interaction and an indirect interaction mediated by the coupler. Moreover, the sign of the second term is different compared to that of the conventional design. As a consequence, the coupler frequency at which $\tilde{g} = 0$ is located below both qubit frequencies. For example, assuming the qubits are in resonance, we obtain $\tilde{g} = 0$ with the coupler frequency 
\begin{equation}\label{eq:omegacoff}
\omega_{\rm c}^{\rm off} = \frac{\omega_1}{\sqrt{1+\xi}} < \omega_1.
\end{equation}
We emphasize that $\tilde{g}$ gives the coupling strength between the hybridized qubit-coupler states which approximate the eigenstates of the setup accurately only in the limit of $g_{12} = 0$. Furthermore, biasing the coupler at the frequency $\omega_{\rm c} = \omega_{\rm c}^{\rm off}$, does not guarantee that the diagonal, Kerr-type couplings are also zero. Especially, the residual $ZZ$ coupling is, in general, not zero at $\omega_{\rm c} = \omega_{\rm c}^{\rm off}$ which is therefore not an optimal choice as the idling frequency.

The computational states of the effective two-qubit system are defined as eigenstates, or dressed states, of the coupled system consisting of the two transmons and the tunable floating coupler. In practice, the computational two-qubit states $|00\rangle$, $|10\rangle$, $|01\rangle$, and $|11\rangle$ are the dressed states that have the largest overlap with the bare-basis states $|\rm ggg\rangle$, $|\rm egg\rangle$, $|\rm gge\rangle$, and $|\rm ege\rangle$ of the non-interacting part of the Hamiltonian (labeling defined as $|\rm Q1,C,Q2\rangle$), respectively. This provides a unique definition of the computational basis in the dispersive regime.

The computational states collect conditional phase through the interactions between the bare-basis states in the two-excitation manifold $\{|\rm eeg\rangle,|ege\rangle,|gee\rangle,| fgg\rangle,|gfg\rangle,|ggf\rangle\}$. The conditional-phase rate, i.e., the effective $ZZ$ interaction strength, is defined as
\begin{equation}
\label{eq:zzcoup}
\zeta = (\omega_{11} - \omega_{10}) - (\omega_{01} - \omega_{00}),
\end{equation}
where $\omega_{ij} / (2\pi)$ are the eigenfrequencies of the computational states. Similar to the effective coupling $\tilde{g}$ in the single-excitation manifold, the conditional-phase rate consists of competing contributions arising from direct and indirect interactions between two-excitation states with different signs. At suitable values for qubit and coupler frequencies, these contributions can cancel each other out, resulting in a negligible effective $ZZ$ interaction rate. We define the idling frequency as the coupler frequency $\omega_{\rm c} = \omega_{\rm c}^{\rm idle}$ for which the effective $ZZ$ interaction is minimized. 

In the dispersive regime, one can compute the conditional-phase rate $\zeta$ analytically using the Schrieffer--Wolff approach, similar to the case of the effective transverse coupling frequency. However, here one needs to do the Schrieffer--Wolff expansion at least to the fourth order in the coupling strengths in order to obtain accurate results. This has been carried out in Refs.~\cite{sung2021realization} and \cite{chu2021}. In our case, however, we often operate in the non-dispersive parameter regime, particularly, close to the resonance between the computational state $|11\rangle$ and non-computational state $|02\rangle$. Consequently, we compute the conditional-phase rate by numerically solving the eigenproblem for the Hamiltonian in Eq.~\eqref{eq:effHam} using $N=3$ states for each transmon in the system.

We show the numerically obtained effective $ZZ$ interaction strength in Fig.~\ref{fig:fig_2}(a) of the main text. Similar to the analytic expression~\cite{sung2021realization,chu2021}, the data show that for the qubit-qubit detunings $\Delta = \omega_1-\omega_2 \in [\alpha_2,-\alpha_1]$, the system has two coupler frequencies for which $\zeta = 0$. In the plane spanned by the coupler frequency and the detuning, the zero-coupling condition $\zeta = 0$ forms a characteristic oval-shaped contour, each point of which can be used as the idling configuration for two-qubit gates.

\section{SIMULATION ON COUPLING STRENGTH FOR DIFFERENT QUBIT-QUBIT DISTANCES}
\label{sec:simulation_coupling_strength}

\begin{table}[tb]
    \centering
    \renewcommand{\arraystretch}{1.4}
    \begin{tabular}{|c|c|}
        \multicolumn{2}{c}{Fig.~\ref{fig:schematic_floating_coupler}} \\
        \hline
        Capacitance & Value (fF)  \\ \hline
      $C_{\rm AB}$ & 93 \\
      $C_{\rm BC}$ & 12 \\
      $C_{\rm CD}$ & 189 \\
      $C_{\rm CF}$ & 41 \\
      $C_{\rm CE}$ & 2 \\
      $C_{\rm sC}$ & 50 \\
      $C_{\rm DE}$ & 6 \\
      $C_{\rm DF}$ & 5 \\
      $C_{\rm sD}$ & 62 \\
      $C_{\rm EF}$ & 192 \\
      $C_{\rm sE}$ & 62 \\
      $C_{\rm FG}$ & 12 \\
      $C_{\rm sF}$ & 47 \\
      $C_{\rm GH}$ & 93 \\
    \hline
    \end{tabular}
    \quad
    \quad
    \begin{tabular}{|c|c|}
        \multicolumn{2}{c}{Fig.~\ref{fig:reduced_floating_coupler}} \\
        \hline
        Capacitance & Value (fF)  \\ \hline
      $C_1$ & 82 \\
      $C_2$ & 82 \\
      $C_{\rm c}$ & 28  \\
      $C_{\rm 1c}^{||}$ & 8 \\
      $C_{\rm 1c}^{\perp}$ & 1 \\
      $C_{\rm 2c}^{||}$ & 8 \\
      $C_{\rm 2c}^{\perp}$ & 1 \\
      $C_{12}$ & 0.07 \\
      $\tilde{C}_{\rm sD}$ & 100 \\
      $\tilde{C}_{\rm sE}$ & 98 \\
    \hline
    \end{tabular}
    \quad
    \quad
    \begin{tabular}{|c|c|}
        \multicolumn{2}{c}{Fig.~\ref{fig:reduced_effective_coupler}} \\
        \hline
        Capacitance & Value (fF)  \\ \hline
      $\bar{C}_{1}$ & 82 \\
      $\bar{C}_{2}$ & 82 \\
      $\bar{C}_{\rm c}$ & 29 \\
      $\bar{C}_{\rm 1c}^{||}$ & 7  \\
      $\bar{C}_{\rm 2c}^{||}$ & 7 \\
      $\bar{C}_{12}$ & 0.25 \\
    \hline
        \multicolumn{2}{c}{Eq.~\eqref{eq:effCapMat}} \\
        \hline
        Capacitance & Value (fF)  \\ \hline
      $C_{\Sigma_1}$ & 91 \\
      $C_{\Sigma_2}$ & 91 \\
      $C_{\Sigma_{\rm c}}$ & 82 \\
      $C_{\rm 1c}$ & 3\\
      $C_{\rm 2c}$ & 3\\
      $C_{\rm 12}^*$ & 0.5\\
    \hline
    \end{tabular}
    \quad
    \quad
    \begin{tabular}{|c|c|}
        \hline
        Josephson energy & Value (GHz)  \\ \hline
      $E_{\rm J_1}$ & 12.1 \\
      $E_{\rm J_2}$ & 13.2 \\
      $E_{\rm J_c}$ & 11.8 \\
    \hline
    \end{tabular}
    \caption{Lumped-element capacitance values for models defined in Figs.~\ref{fig:schematic_floating_coupler},~\ref{fig:reduced_floating_coupler}, and~\ref{fig:reduced_effective_coupler}. We also show the elements of the effective capacitance matrix in Eq.~\eqref{eq:effCapMat}, and the designed values for Josephson energies used in the models.}
    \label{tab:capacitances}
\end{table}

The sample is fully designed with KQCircuits~\cite{kqcircuits}, our open source Python package for superconducting circuit design, from which we can export netlists. We model the two-qubit system as a quasi-lumped-element network model using scikit-rf~\cite{scikit}. In this model, the qubits and the coupler are represented by lumped-element circuits, whereas the waveguide extenders are modeled as transmission lines with analytical expressions for the effective dielectric constant and the characteristic impedance. We fix the capacitances of the system as shown in Table~\ref{tab:capacitances} and use Ansys Q3D Extractor to find the shapes of the qubits, the coupler, and the waveguide extender capacitor pads which realize the required capacitance values.

To study the coupling strength as a function of the qubit spacing, as given in Fig.~\ref{fig:fig_1}(b) of the main text, we change the length of the waveguides in the quasi-lumped-element model, while keeping the geometry of the finger and gap capacitors the same. We then reduce the network to the one given in Fig.~\ref{fig:reduced_effective_coupler} and calculate the effective capacitances from the admittance matrix of the network at \SI{5}{\giga\hertz}. Lastly, we use Eq.~\eqref{eq:coupling_strengths} to calculate the coupling strengths shown in Fig.~\ref{fig:fig_1}(b) in the main text. Here, we use the Josephson energies shown in Table~\ref{tab:capacitances} to obtain the frequencies $\omega_1/(2\pi) = \SI{4.10}{\giga\hertz}$, $\omega_2/(2\pi) = \SI{3.89}{\giga\hertz}$, and $\omega_{\rm c}/(2\pi) = \SI{3.195}{\giga\hertz}$ for $\rm{Q}_1$, $\rm{Q}_2$, and the coupler, respectively.

\section{SIMULATION ON CROSSTALK OF PASSING TRANSMISSION LINE}
\label{sec:simulation_crosstalk}

In this section, we will elaborate on the details of the simulation results, where we simulate the crosstalk between a transmission line running \SI{8}{\micro\meter} above the qubit-coupler-qubit structure and each component, as shown in Fig.~\ref{fig:fig_1}(d) of the main text. In this simulation, a \SI{3.5}{\milli\meter}-long transmission line on the top chip crosses the qubit-coupler-qubit line on the bottom chip almost perpendicularly. The line is long compared to our qubit size of \SI{0.5}{\milli\meter} to account for all of the extra capacitance. We simulate the sample with finite element simulation, where the other capacitances are listed in Table~\ref{tab:capacitances}. We place five ports on the simulated sample: one at each qubit ($\rm{Q}_1$ and $\rm{Q}_2$), one at each island of the coupler ($\rm{c}_C$ and $\rm{c}_F$) and one at the end of the transmission line. We extract the capacitance to the crossing transmission line from the obtained admittance matrix at \SI{4.3}{\giga\hertz}. This frequency is chosen to represent the worst case, as it is at an identical frequency to the designed qubit frequencies. Since the coupler is floating, we are interested in the differential capacitance between the two coupler islands $C_{\rm{TC}\leftrightarrow \rm{TL}} = |C_{\rm{c_C}\leftrightarrow \rm{TL}} - C_{\rm{c_F}\leftrightarrow \rm{TL}} |$. As a consequence, there is a crossing position of the transmission line, where $C_{\rm{c_C}\leftrightarrow \rm{TL}} = C_{\rm{c_F}\leftrightarrow \rm{TL}}$. Since our coupler structure is symmetric around its center, a dip in the coupling ratio $r_{\rm c}$ of the coupler at $x_{\rm cross}=0$ appears in Fig.~\ref{fig:fig_1}(d) in the main text.

To confirm that the excitation of \SI{4.3}{\giga\hertz} on the transmission line does not excite any particular mode, we repeat the simulations for different frequencies in the range between \SI{1}{\giga\hertz} to \SI{10}{\giga\hertz} and find no quantitative deviations from the simulations with a \SI{4.3}{\giga\hertz} excitation.

\section{EXPERIMENTAL SETUP}
\label{sec:experimental_setup}

\begin{figure}
    \centering
    \includegraphics[width=1.0\columnwidth]{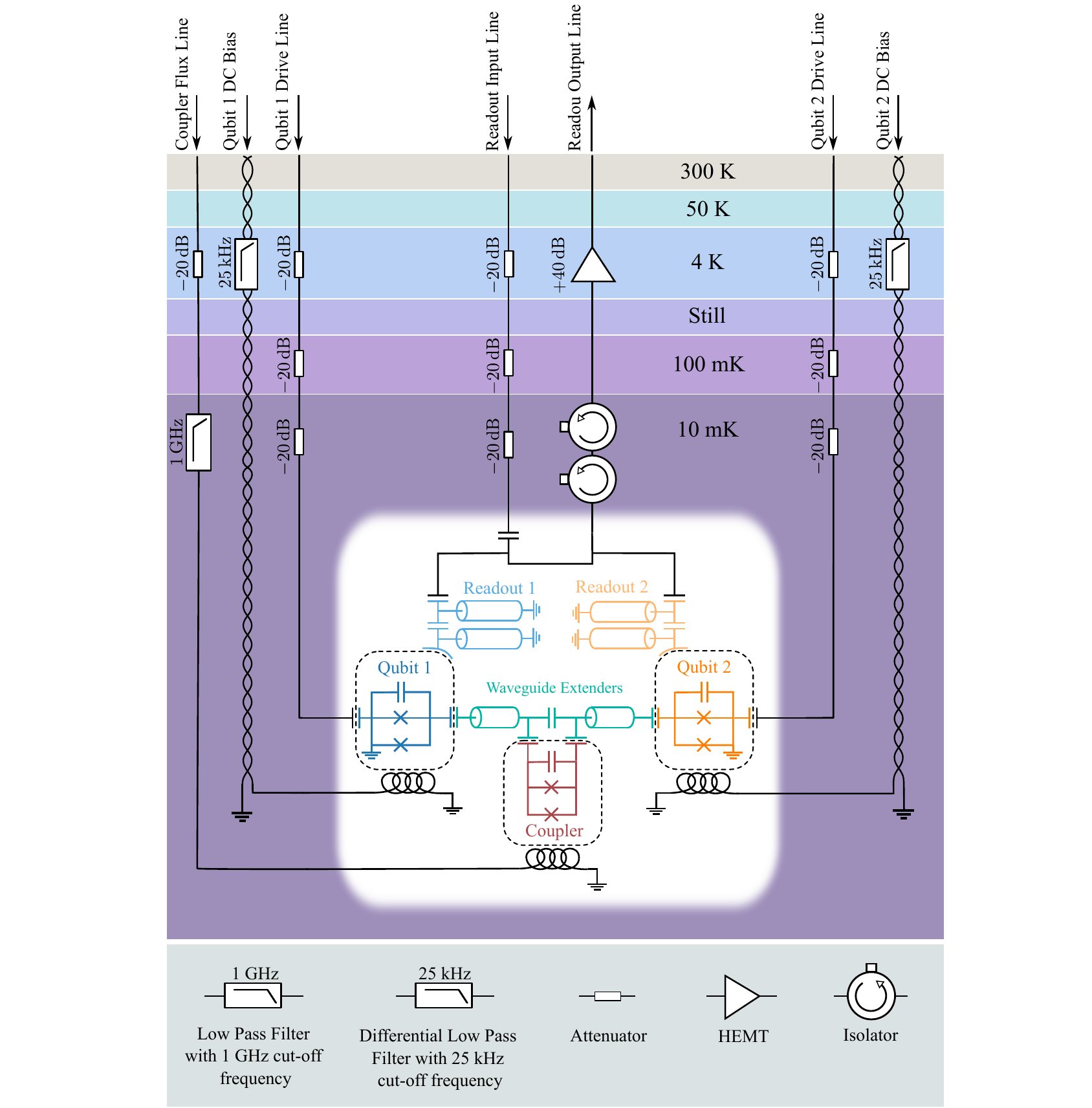}
    \caption{The diagram of the used experimental setup. The background colors depict the different temperature stages of the dilution refrigerator. The coaxial microwave cables (straight black lines) and twisted-pair cables (braided lines) are used to control and read out the qubit and coupler states. The schematic of the sample in the white box consists of two flux-tunable qubits (blue and orange), a readout structure for each qubit (light blue and light orange), and a flux-tunable coupler (red).}
    \label{fig:cryo_setup}
\end{figure}

The used experimental setup including a schematic of the sample is shown in Fig.~\ref{fig:cryo_setup}. Both qubits used in the experiment are flux-tunable transmon qubits. Single-qubit gates are implemented using capacitively coupled drive lines attenuated by a total of \SI{60}{\dB} throughout the different temperature stages. Each of the \SI{20}{\dB} attenuators are thermalized at the corresponding temperature stages, as shown in Fig.~\ref{fig:cryo_setup}. The pulse envelopes for the qubit drives are generated by a Zurich Instrument arbitrary-waveform generator HDAWG and the carrier for the in-phase and quadrature (IQ) mixing to the qubit frequency via the Analog Devices mixer HMC8193 is provided by a Rohde\&Schwarz microwave generator SGS100A. To control the frequency of each qubit, a voltage source at room temperature is connected to a twisted-pair cable which reaches down to the \SI{10}{\milli\kelvin} stage. One conductor of the twisted pair is grounded near the sample and the other conductor leads through the sample as an on-chip flux line and is then grounded on the sample near the SQUID loop. To filter the noise from the voltage source and the noise picked up by the twisted pair, we employ a \SI{25}{\kilo\hertz} second-order differential low-pass filter thermalized at the \SI{4}{\kelvin} stage. 

\begin{table}[tb]
    \centering
    \renewcommand{\arraystretch}{1.7}
    \begin{tabular}{m{0.6cm} >{\centering\arraybackslash} m{1.0cm}  >{\centering\arraybackslash} m{1.1cm}  >{\centering\arraybackslash} m{1.0cm}  >{\centering\arraybackslash} m{1.0cm}  >{\centering\arraybackslash} m{1.0cm} }
        & & \multicolumn{4}{ c }{Measured State} \\
        & \multicolumn{1}{c|}{}& $\ket{\rm 00}$ & $\ket{\rm 01}$ & $\ket{\rm 10}$ & $\ket{\rm 11}$\\ \cline{2-6}
        \multicolumn{1}{c}{\multirow{4}{*}{\rotatebox[origin=c]{90}{Prepared state}}}
        & \multicolumn{1}{c |}{$\ket{\rm 00}$} & 0.902 & 0.055 & 0.040 & \multicolumn{1}{c}{0.003} \\ 
        & \multicolumn{1}{c |}{$\ket{\rm 01}$} & 0.052 & 0.905 & 0.003 & \multicolumn{1}{c}{0.040} \\
        & \multicolumn{1}{c |}{$\ket{\rm 10}$} & 0.065 & 0.004 & 0.880 & \multicolumn{1}{c}{0.050} \\
        & \multicolumn{1}{c |}{$\ket{\rm 11}$} & 0.004 & 0.061 & 0.110 & \multicolumn{1}{c}{0.824} \\
    \end{tabular}
    \caption{Readout correlation matrix for the two-qubit system which is prepared $2\times 10^4$ times in each state in the computational basis $\{\ket{\rm 00}, \ket{\rm 01}, \ket{\rm 10}, \ket{\rm 11}\}$. The single-shot measurement probabilities are then obtained for each of the four computational states using previously determined discrimination lines in the in-phase--quadrature-phase plane of the measurement signal of each qubit.}
    \label{tab:readout_fidelities}
\end{table}

Each qubit has its own readout structure which consists of a readout resonator and a Purcell filter \cite{Sete2015, heinsoo2018rapid}. Both readout structures are coupled to a single transmission line. The signal traveling to the output line is amplified with a low-noise high-electron-mobility transistor (HEMT) at the \SI{4}{\kelvin} stage. Despite not having a quantum-limited amplifier in the setup, readout fidelities around \SI{88}{\percent} are reached due to strongly coupled readout resonators designed for high fidelity readout, see Table~\ref{tab:readout_fidelities} for details. The probe pulse at the intermediate frequency is generated and the response is digitized using a Zurich Instrument Quantum Analyzer UHFQA. For up and down conversion to the readout frequency, a single Rohde\&Schwarz microwave generator is used as a local oscillator.

The tunable-coupler structure includes a flux-tunable floating transmon qubit. Since the coupler state does not necessarily need to be read out, no separate readout structure was designed for the coupler. However, by swapping the coupler state to the qubit or by using coupler-state-dependent dispersive shift of a qubit, it is still possible to read out the state of the coupler for characterization experiments, see Appendix~\ref{sec:coupler_characterization}. On this device, the tunable coupler is the only component which needs fast flux control. For improved coupler coherence, a \SI{1}{\giga\hertz} low-pass filter is installed to the flux line at the \SI{10}{\milli\kelvin} stage of the cryostat. The direct current (DC) bias of the coupler flux line and the fast pulses are generated by a single channel of a Zurich Instrument arbitrary-waveform generator.

The summary of the device parameters is provided in Table~\ref{tab:parameters}. Here, the frequencies and the coherence times are given both at the flux insensitive point (sweet spot) and at the two-qubit-gate idling configuration. The two-qubit-gate idling point is chosen such that the frequency of the $\ket{11}$ state is close to the $\ket{02}$ state to enable fast CZ gates. However, since we chose to not have fast flux control for the qubits, the smaller the detuning between the $\ket{11}$ and $\ket{02}$ states, the larger the effect of drive crosstalk. In particular, when driving the $\ket{0} \leftrightarrow \ket{1}$ transition of $\rm{Q}_1$, we also spuriously drive the $\ket{1} \leftrightarrow \ket{2}$ transition of $\rm{Q}_2$. Therefore, we tune the qubit frequencies to the values shown in Table~\ref{tab:parameters} to have an ideal trade-off between fast two-qubit gates and weak drive crosstalk.

\begin{table}[tb]
    \centering
    \renewcommand{\arraystretch}{1.4}
    \begin{tabular}{|l| C C C|}
        \hline
         Quantity, symbol (unit) & Qubit 1 & Qubit 2 & Coupler  \\ \hline
        Readout resonator frequency, $\omega_\mathrm{R}/2\pi$ (GHz)& 4.950 & 6.134 & -\\
        Readout Purcell filter frequency, $\omega_\mathrm{Rpf}/2\pi$ (GHz) & 5.037 & 6.125 & -\\
        Effective readout resonator bandwidth, $\kappa_\mathrm{eff} / 2\pi$ (MHz) & 13 & 22 & -\\
        Readout circuit dispersive shift, $\chi_\mathrm{R} / 2 \pi$ (MHz) & 5.3 & 4.7 & -\\
        Qubit frequency at sweet spot, $\omega_\mathrm{ge}/2\pi$ (GHz) & 4.102 & 3.972 & 4.210\\
        Qubit frequency at 2QG idling point, $\omega_\mathrm{ge}/2\pi$ (GHz) & 4.102 & 3.892 & 3.195\\
        Transmon anharmonicity, $\alpha$ (MHz) & $-215$ & $-217$ & $-250$\\
        Energy relaxation time at sweet spot, $T_1$ (µs) & 13 & 39 & 30\\
        Energy relaxation time at 2QG idling point, $T_1$ (µs) & 13 & 42 & 50\\
        Transverse relaxation time at Sweet spot, $T_2$ (µs) & 14 & 14 & 1.4\\
        Transverse relaxation time at 2QG idling point, $T_2$ (µs) & 14 & 8 & 0.35\\
    \hline
    \end{tabular}
    \caption{Summary of device parameters. Parameters at the flux insensitive point (sweet spot) and at the two-qubit gate (2QG) idling configurations are listed.}
    \label{tab:parameters}
\end{table}

\section{DEVICE FABRICATION}
\label{sec:sample_fabrication}
The sample used in the experiments is fabricated at the OtaNano Micronova cleanroom. First, a high-purity \SI{200}{\nano\meter}-thick niobium layer is deposited on a high-resistivity ($\rho >\SI{10}{\kilo\ohm \cm}$) non-oxidized $n$-type undoped (100) six-inch silicon wafer by sputtering. The coplanar waveguides and capacitive structures are then formed by photolithography with subsequent reactive ion etching. After etching, the photoresist residuals are cleaned in an ultrasonic bath with acetone and isopropanol. Next, electron beam lithography is used with subsequent electron beam shadow evaporation and lift-off processes to form the Josephson junctions of the qubits. The qubit junctions are formed by two \SI{20}{\nano\meter}-thick aluminum layers. Before the evaporation of these structures, natural oxides are removed from the surface by argon ion milling. Finally, \SI{800}{\nano\meter}-thick aluminum airbridges are fabricated by sputtering and subsequent lift off.
After the device has been fabricated, we measure the room temperature resistance of the qubit junctions, producing the scratches on the qubit islands seen in Fig.~\ref{fig:fig_1}(b) in the main text.

\section{COUPLER CHARACTERIZATION}
\label{sec:coupler_characterization}
\subsection{Three-tone spectroscopy} 

In the absence of a dedicated readout circuit for the coupler, we use three-tone spectroscopy with $\rm{Q}_2$ as an ancilla qubit to find the coupler frequency~\cite{li2020tunable}.
We begin by applying a weak probe tone to $\rm{Q}_2$ at its transition frequency  (see the experimental setup in Fig.~\ref{fig:cryo_setup} and the qubit properties in Table~\ref{tab:parameters}) and monitor its state by sending a continuous wave to its readout resonator. At the same time, we strongly drive the coupler through the drive line of $\rm{Q}_1$. When the coupler drive tone is in resonance with the coupler frequency, the coupler gets partially excited, and $\rm{Q}_2$ frequency shifts down due to dispersive interaction between the qubit and the coupler. As a result, the weak probe tone driving $\rm{Q}_2$ no longer excites $\rm{Q}_2$, leading to a change in the readout signal of $\rm{Q}_2$. This effectively maps the coupler state to the readout signal of $\rm{Q}_2$.

\begin{figure*}
    \centering
    \includegraphics[width=0.5\columnwidth]{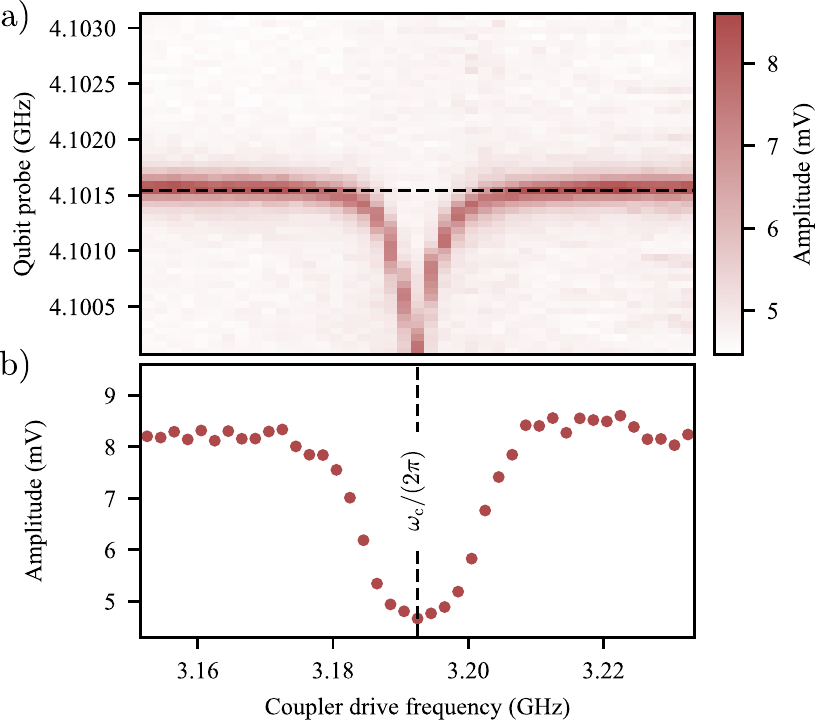}
    \caption{Results from a three-tone spectroscopy experiment. (a) Qubit spectroscopy with an additional frequency sweep of the coupler drive. The horizontal black dashed line indicates the one-dimensional sweep for a three-tone experiment shown in (b). The coupler frequency $\omega_{\rm c}/ (2 \pi)$ is extracted by a minimum search of the one-dimensional sweep (vertical black dashed line).
    }
    \label{fig:three_tone_sweep_and_freqs}
\end{figure*}

To illustrate the effect of the qubit probe frequency and the coupler drive frequency, we sweep both of the parameters and plot the readout voltage, see Fig.~\ref{fig:three_tone_sweep_and_freqs}(a). When the coupler drive frequency is far away from the coupler transition frequency of $\omega_{\rm c}/ (2 \pi) = \SI{3.192}{\giga\hertz}$, the experiment resembles a normal qubit spectroscopy experiment with the qubit frequency at \SI{4.1016}{\giga\hertz}. As the coupler drive frequency gets close to the coupler transition frequency, the frequency of $\rm{Q}_2$ decreases down to \SI{4.100}{\giga\hertz} due to dispersive shift. By fixing the probe tone at the frequency of $\rm{Q}_2$ and sweeping the coupler drive frequency, we implement the coupler frequency spectroscopy experiment, see Fig.~\ref{fig:three_tone_sweep_and_freqs}(b). In Fig.~\ref{fig:fig_2}(c) of the main text, this method is used for measuring coupler frequencies below \SI{3.8}{\giga\hertz}, where the coupling between the hybridized coupler and qubit mode to the readout resonators is too weak for standard dispersive readout.

Although this method works with either $\rm{Q}_1$ or $\rm{Q}_2$ as the ancilla, the transition frequency difference between $\ket{000} \leftrightarrow \ket{001}$ and $\ket{010} \leftrightarrow \ket{011}$ with $\rm{Q}_2$ as an ancilla is larger than with $\rm{Q}_1$ as an ancilla.

\subsection{Coupler coherence and flux noise}

\begin{figure*}
    \centering
    \includegraphics{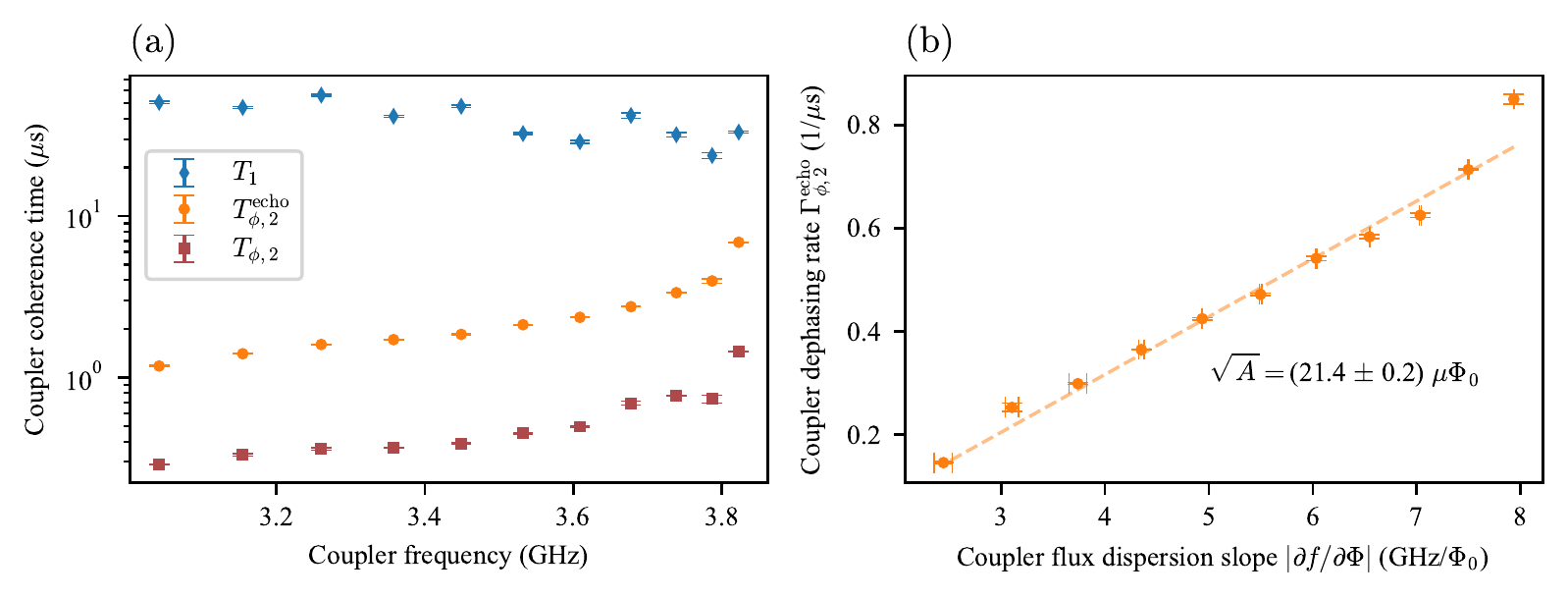}
    \caption{(a) Experimentally obtained relaxation and coherence times of the coupler as functions of its frequency. (b) Pure dephasing rate from echo measurements as a function of the flux dispersion slope. A line fit yields the flux noise amplitude $\sqrt{A}$.}
    \label{fig:coupler_coherence_flux_noise}
\end{figure*}

Above, we discussed the three-tone spectroscopy measurement in the context of continuous waves and observed that it allows us to read out the coupler state.
Here, we use two-probe readout in standard time-dependent experiments to characterize the coupler coherence as a function of the coupler frequency, see Fig.~\ref{fig:coupler_coherence_flux_noise}(a) for the results. Instead of a continuous drive, the ancilla qubit and the readout resonator are now pulsed consecutively. The length of the $\pi$-pulse on the ancilla qubit is chosen to have the spectral width narrower than the coupler state-dependent frequency shift. The dephasing rates for the echo and Ramsey experiments are calculated from the fit to Eq. \eqref{eq:decay_envelope} in the main text, where $T_1$ is acquired from the energy relaxation measurements of the coupler.

We can utilize the extracted data for $T_{\phi, 2}^{\rm echo}$ to characterize the $1/f$ flux noise of the coupler \cite{Ithier2005, yoshihara2006decoherence, braumuller2020characterizing, bylander2011noise}.
In the presence of $1/f$ flux noise, the pure dephasing rate is
\begin{align}
    \Gamma_{\phi, 2}^{\rm echo} = 1/T^{\rm echo}_{\phi, 2} = \sqrt{A \ln 2} \left \vert \frac{\partial\omega}{\partial\Phi} \right\vert, \label{eq:dephasing_vs_slope}
\end{align}
where $\left \vert \partial\omega/\partial\Phi \right\vert$ is the slope of the coupler flux dispersion curve and $\sqrt{A}$ is the flux noise amplitude.
The power spectral density can then be extracted as $S(\omega) = A/\vert\omega\vert$, assuming $1/f$ flux noise \cite{Ithier2005, yoshihara2006decoherence, bylander2011noise, braumuller2020characterizing}.
In Fig.~\ref{fig:coupler_coherence_flux_noise}(b), the pure dephasing rate $\Gamma_{\phi, 2}^{\rm echo}$ is shown as a function of the flux dispersion slope $\left \vert \partial f/\partial\Phi \right\vert = (2\pi)^{-1}\left \vert \partial\omega/\partial\Phi \right\vert$.
Each flux dispersion slope point is extracted by fitting a cubic polynomial to the measured flux dependency of the coupler frequency and then extracting the derivative.
We fit Eq.~\eqref{eq:dephasing_vs_slope} together with an offset parameter to the data, resulting in a flux noise amplitude of $\sqrt{A} = \SI[separate-uncertainty = true]{21.4(2)}{\micro\Phi_0}$.
From the fit, we extract an offset of \SI[separate-uncertainty = true]{-0.132(6)}{\micro\second^{-1}}, where the deviation from zero is possibly caused by other high-frequency dephasing processes~\cite{braumuller2020characterizing}.

We compare the extracted flux noise amplitude to the theoretically expected value given the geometry of the coupler SQUID loop following the analysis of Ref.~\cite{braumuller2020characterizing}.
From the geometry, we estimate $\sqrt{A} = \SI{7.2}{\micro\Phi_0}$, which is significantly smaller than the measured value. The source of the excess flux noise unexplained by the SQUID geometry is unknown, but might be originating from the flux noise of the voltage source.
Furthermore, we can compare the coupler flux noise amplitude to the measured qubit flux noise amplitudes which are $\SI[separate-uncertainty = true]{4.12 \pm 0.04}{\micro\Phi_0}$ and $\SI[separate-uncertainty = true]{4.29\pm 0.05}{\micro\Phi_0}$ for $\rm Q_1$ and $\rm Q_2$, respectively.
These values correspond well to the expected value from the SQUID geometry alone ($\SI{4.9}{\micro\Phi_0}$).
Despite the coupler suffering from excess flux noise, the dephasing time of the coupler is not expected to limit our gate fidelity, as we have shown in Fig.~\ref{fig:fig_4}(d) in the main text.

\section{RANDOMIZED BENCHMARKING WITH MULTIPLE INTERLEAVED CZ GATES}
\label{sec:ncz_irb}
\begin{figure*}
    \centering
    \includegraphics[width=0.5\columnwidth]{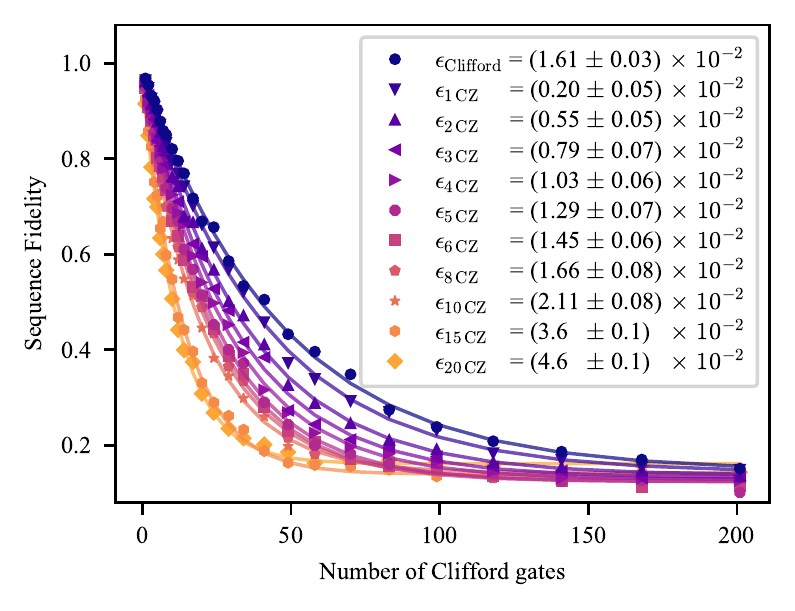}
    \caption{Experimentally obtained sequence fidelity as a function of the number of Clifford gates in the randomized benchmarking sequence with $n$ interleaved CZ gates, where $n$ ranges from 1 to 20. Each data point (symbols) represents the average result of 30 random sequences. Each sequence fidelity trace is fitted using an exponential model (solid lines) to extract the error $\epsilon_{n \rm CZ}$ per $n$ CZ gates.}
    \label{fig:SM:ncz_irb}
\end{figure*}

To further illustrate the results obtained in Fig.~\ref{fig:fig_3}(b), we show in Fig.~\ref{fig:SM:ncz_irb} the sequence fidelity of the randomized benchmarking for each number $n$ of interleaved CZ gates. We fit the sequence fidelity of the reference trace $\mathcal{F}_{\rm ref}$ with the exponential model $\mathcal{F}_{\rm ref} = A {p_{\rm ref}}^{m} + B$, where $p_{\rm ref}$ is the sequence decay, $m$ is the number of Clifford gates, and $A$ and $B$ are parameters to capture the state preparation and measurement errors \cite{barends2014superconducting}. The error per Clifford is given by $\epsilon_{\rm Clifford} = (1 - p_{\rm ref}) (d - 1)/d $, where $d=2^{ N_{\rm q}} = 4$ is the dimensionality of the hilbert space for $N_{\rm q}$ qubits. This model can be applied to fit the sequence fidelity $\mathcal{F}_{n\rm CZ}$ for $n$ interleaved CZ gates and extract the sequence decay $p_{n\rm CZ}$. The errors per $n$ CZ gates can then be calculated as $\epsilon_{n \rm CZ} = (1 - p_{n \rm CZ}/p_{\rm ref}) (d - 1)/d $, where the experimental results are shown in the legend of Fig.~\ref{fig:SM:ncz_irb}. The error per CZ gate $\epsilon_{\rm CZ}^{(n)}$ can then be calculated as $\epsilon_{\rm CZ}^{(n)} = 1 - \sqrt[\leftroot{-3}\uproot{3}n]{1 - \epsilon_{n \rm CZ}}$ to reach the results shown in Fig.~\ref{fig:fig_3}(b) of the main text. For small $\epsilon_{n \rm CZ}$, this expression reduces to $\epsilon_{\rm CZ}^{(n)} \approx \epsilon_{n \rm CZ} / n$. Here we assume that each of the $n$ CZ gates contributes with an equal amount to the total error $\epsilon_{n \rm CZ}$. This assumption becomes invalid as soon as the distortion of one flux pulse affects one or more of its subsequent flux pulses. Therefore, it is important to compensate for the flux pulse distortions in order for this assumption to hold.

\section{ESTIMATING COHERENCE LIMIT OF AVERAGE GATE ERRORS}
\label{sec:coherence}

We estimate the coherence limit of the gate error measured with randomized benchmarking (RB) using the theoretical results derived in Refs.~\cite{chu2021, abad2021universal}. We start by writing the coherence limit for two isolated qubits, and later generalize the result for the qubit-coupler-qubit system using the effective coherence time method presented in Refs.~\cite{xu2020high, chu2021}.

\subsection{Coherence limit for CZ gate}

For simplicity, we assume that the noise sources acting on the qubits are independent, which is mostly accurate for the dominating noise sources in our system: longitudinal noise and flux noise from the oxygen impurities in the SQUID loops~\cite{braumuller2020characterizing}. We also assume that the environment of the qubits is cold, which allows us to neglect the excitation rate of the qubits. Then, in the limit of small errors and Markovian noise, the error rate from energy-relaxation can be written as \cite{abad2021universal}
\begin{equation}
\label{eq:eps_T1}
    \epsilon_{T_1} = \sum_i^{N_{\rm q}} \frac{N_{\rm q}d}{2(d+1)}\frac{\tau}{T_1^{{\rm Q}_i}},
\end{equation}
where $d = 2^{N_{\rm q}}$ is the dimensionality of the Hilbert space for $N_{\rm q}$ qubits.

As opposed to energy relaxation, which is typically Markovian, the impact of phase noise significantly depends on the temporal correlations in the phase fluctuations, often invalidating the Markov approximation. Nevertheless, the amplitude of phase noise is typically assumed to be Gaussian distributed, with a notable counter-example being if the qubit is operated at or near its flux sweet spot \cite{Sung2019}. However, the non-Gaussian component is often relatively small compared to the dominating noise sources, and thus it can be typically neglected for practical applications. A broad review of the impact of phase noise on the decay of qubit coherence can be found in Ref.~\cite{Ithier2005}. For the sake of simplicity, we only discuss two extreme cases here, quasi-static noise and white noise, and divide all the noise sources into either one of these two categories.

White noise has a uniform noise spectral density over a wide band of frequencies and, therefore, does not have any temporal correlations, enabling Markov approximation. This results in exponential decay in the off-diagonal elements in the qubit density matrix, characterized by the decay envelope

\begin{equation}
    \chi(t) = \mathrm{e}^{-\frac{t}{T_{\phi, 1}}},
\end{equation}
similar to the result obtained with the Lindblad master equation. 

Quasi-static noise arises from any noise source whose noise power originates from frequencies below the smallest relevant rates in the system, typically the repetition rate of the experiment. On the other hand, noise at frequencies below the calibration rate of the system can be considered static and fully eliminated by the calibration. Quasi-static noise therefore covers all the noise power roughly between the repetition rate of the experiment and the calibration rate of the experiment, $P_{\rm q-s} = \int_{1/t_{\rm repetition}}^{1/t_{\rm calibration}} S_{\rm ff}(f) \mathrm{d}f$, where $S_{\rm ff}(f)$ is the single-sided noise power spectral density of the qubit frequency fluctuations. Quasi-static noise results in Gaussian decay envelope for the qubit coherence, $\chi(t) = \mathrm{e}^{-\left(t/T_{\phi, 2}\right)^2}$, where $T_{\phi, 2} = 1/P_{\rm q-s}$ is the dephasing time \cite{Ithier2005}.

Noise with a significant spectral weight in between these two extreme cases results in decay envelopes that complicate the analytical calculations. Most importantly, the dephasing time for $1/f^\alpha$-noise can no longer be simply defined as a single coefficient. For simplicity, here we approximate $1/f^\alpha$ as quasi-static since its decay envelope only significantly deviates from Gaussian when $t$ approaches the root for $\chi(t) = 1/\mathrm{e}$. The most important difference between $1/f^\alpha$ noise and quasi-static noise is observed when noise-decoupling sequences are considered, for example, spin echo completely eliminates quasi-static noise but not $1/f^\alpha$-noise \cite{Ithier2005, bylander2011noise}.

With these assumptions, the contributions from white noise and quasi-static noise to the gate error rate can be modeled as \cite{abad2021universal}
\begin{equation}
\label{eq:eps_Tphi1}
    \epsilon_{T_{\phi, 1}} = \sum_i^d \frac{N_{\rm q}d}{2(d+1)}\frac{\tau}{T_{\phi, 1}^{{\rm Q}_i}},
\end{equation}
and for quasi-static noise as \cite{chu2021}
\begin{equation}
\label{eq:eps_Tphi2}
    \epsilon_{T_{\phi, 2}} = \sum_i^d \frac{N_{\rm q}d}{2(d+1)}\left(\frac{\tau}{T_{\phi, 2}^{{\rm Q}_i}}\right)^2.
\end{equation}

\subsection{Coherence limit for CZ gate with tunable coupler}
As the coupler hybridizes with the qubits during the gate, the coherence of the computational states is affected by the coupler coherence. In Ref.~\cite{xu2020high}, a method to estimate the CZ gate fidelity based on effective qubit coherence times was proposed. The qubit coherence times are measured at each of the coupler frequency relevant for the gate, and an effective coherence time is calculated as an average over the coherence times weighted by the time coupler spends at each of those frequencies.

This yields
\begin{equation}
\label{eq:T_1_eff}
    T_1^{\rm eff} = \tau/\int_0^\tau 1/T_1[\omega_{\rm c}(t)] \mathrm{d}t,
\end{equation}

\begin{equation}
\label{eq:T_phi_white_eff}
    T_{\phi, {\rm 1}}^{\rm eff} = \tau / \int_0^\tau 1/T_{\phi, {\rm 1}}[\omega_{\rm c}(t)] \mathrm{d}t,
\end{equation}

\begin{equation}
\label{eq:T_phi_qs_eff}
    T_{\phi, {\rm 2}}^{\rm eff} = 1/\sqrt{\frac{1}{\tau}\int_0^\tau \left\{1/T_{\phi, {\rm 2}}[\omega_{\rm c}(t)]\right\}^2 \mathrm{d}t},
\end{equation}
where the weighting is carried out in terms of decoherence rates, and $\omega_c(t)$ is the coupler pulse shape. Additionally, the quasi-static dephasing rates need to be averaged under a square root due to their quadratic contribution to the error rates. In the above analysis, we have assumed that both qubits contribute equally to the CZ gate error rate. However, during the gate the state $\ket{11}$ evolves to $\ket{02}$ and back, reducing the probability of a relaxation error for $Q_1$ and correspondingly increasing it for $Q_2$.

\begin{figure}[tb]
    \centering
    \includegraphics[width=0.5\columnwidth]{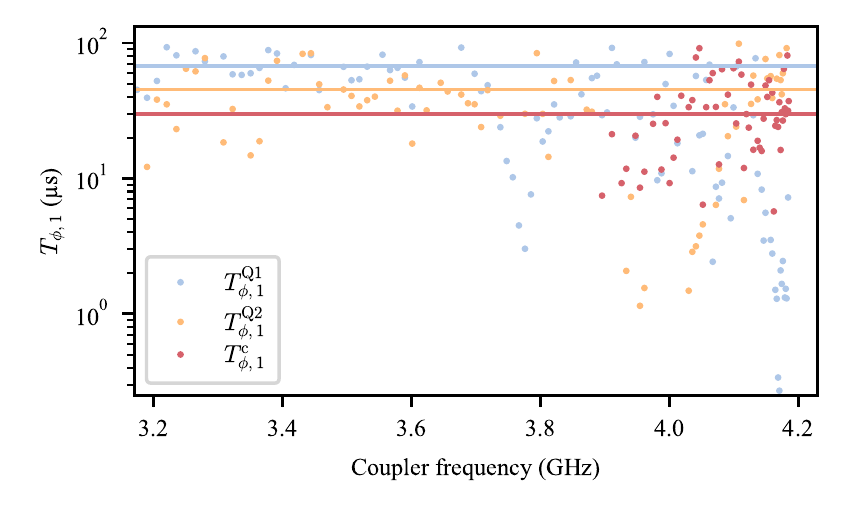}
    \caption{Figure shows the exponential part of the dephasing times as a function of the coupler frequency for the hybridized modes. For the coupler frequencies relevant for the CZ gate, the exponential part of dephasing mostly stays constant, except for a sharp drop at \SI{3.8}{\giga\hertz} for qubit 1. We observed the feature to move to different frequencies over time indicating that it originates from a spectrally active TLS. The solid lines show the median of the measured exponential dephasing rates, evaluated at the frequencies relevant for the CZ gate, and used for calculating $T_{\phi, 1, {\rm eff}}^k$.}
    \label{fig:T_phi_exp}
\end{figure}

To model the effective decoherence rates based on the decoherence rates of the uncoupled modes, we calculate the participation ratios of the uncoupled modes in the computational states for various coupler frequencies, see Appendix~\ref{sec:effective_coupling} for details. The energy-relaxation rate of the coupled mode is then
\begin{equation}
    \Gamma_1^k = 1/T^k_1 = \sum_{i={\rm Q_1}, {\rm Q_2}, {\rm c}} p_{i, k}\tilde{\Gamma}^i_1,
\end{equation}
where $\tilde{\Gamma}^i_1$ is the energy-relaxation rate of the uncoupled mode $i$ and $p_{i, k}$ is the participation ratio of uncoupled mode $i$ in the coupled mode $k$. We note that due to the hybridization of the modes, the energy-relaxation events become correlated with each other even though the energy-relaxation events of the individual components were assumed to be independent. Nevertheless, for first-order Markovian noise processes resulting from a linear coupling to a bath, the error rate is identical for correlated and non-correlated noise \cite{abad2021universal}, allowing us to use Eq.~\eqref{eq:eps_T1} to calculate the expected error. Similarly, we can calculate the dephasing rate corresponding to the Markovian part of the dephasing noise as

\begin{equation}
    \Gamma_{\phi, 1}^k = 1/T^k_{\phi, 1} = \sum_{i={\rm Q_1}, {\rm Q_2}, {\rm c}} p_{i, k}\tilde{\Gamma}^i_{\phi, 1},
\end{equation}
where, as before, $\tilde{\Gamma}^i_{\phi, 1}$ is the exponential dephasing rate of the uncoupled mode $i$ and $p_{i, k}$ is the participation ratio. Fig. \ref{fig:T_phi_exp} shows the measured values for $T^i_{\phi, 1}$ and the solid line which indicates the value for $T^i_{\phi, 1, {\rm eff}}$ used in the main text.

To evaluate the Gaussian part of the dephasing noise for the coupled mode, we need to separately account for the independent low-frequency fluctuations in the SQUID loops of the qubits and the couplers. The combined effect of the fluctuations on the dephasing of the coupled mode is \cite{chu2021, Ithier2005}
\begin{equation}
\label{eq:Gamma_phi_2}
    \Gamma_{\phi, 2}^{k} = 1/T_{\phi, 2}^{k} =\sqrt{\sum_{i=\rm{Q_1}, \rm{Q_2}, \rm{c}} \left(p_{i, k}\tilde{\Gamma}_{\phi, 2}^i\right)^2},
\end{equation}
where $p_{i, k}$ is the participation ratio of the uncoupled mode $i$ in the hybridized mode $k$ and $\tilde{\Gamma}_{\phi, 2}^i$ are the dephasing rates of the uncoupled modes. Strictly speaking, Eq. \eqref{eq:eps_Tphi1} is no longer valid for the non-Markovian correlated noise described by Eq. \eqref{eq:Gamma_phi_2}, but for simplicity, we decide to rely on it for the estimate of the coherent error limit.

\end{document}